\documentclass[aps, preprint,superscriptaddress,floatfix]{revtex4-2}
\usepackage{graphicx} 
\usepackage{subcaption}
\usepackage{amsmath}
\usepackage{booktabs}
\usepackage{xcolor}
\usepackage[version=4]{mhchem}
\usepackage{indentfirst}

\usepackage{float}

\begin{document}

\title{Stability and Dynamics of Sn-based Halide Perovskites: Insights from MACE-MP-0 and Molecular Dynamics Simulations}
\author{Thiago Puccinelli}
\affiliation{Departamento de F\'isica dos Materiais e Mec\^anica, Instituto de F\'isica, Universidade de S\~ao Paulo, S\~ao Paulo 05508-090, S\~ao Paulo, Brazil}

\author{Lucas Martin Farigliano}
\affiliation{Departamento de F\'isica dos Materiais e Mec\^anica, Instituto de F\'isica, Universidade de S\~ao Paulo, S\~ao Paulo 05508-090, S\~ao Paulo, Brazil}
\affiliation{INFIQC, CONICET, Departamento de Qu\'imica Te\'orica y Computacional, Facultad de Ciencias Qu\'imicas, Universidad Nacional de C\'ordoba, Argentina}

\author{Gustavo Martini Dalpian}
\affiliation{Departamento de F\'isica dos Materiais e Mec\^anica, Instituto de F\'isica, Universidade de S\~ao Paulo, S\~ao Paulo 05508-090, S\~ao Paulo, Brazil}

\begin{abstract}

Tin-based halide perovskites have emerged as promising lead-free alternatives for optoelectronic applications, yet their structural stability and phase behavior at finite temperatures remain challenging to predict. Here, we assess the predictive capabilities of the foundational machine learning model MACE-MP-0 — trained on a broad chemical space and applied without system-specific fine-tuning — for the temperature-dependent behavior of \ce{CsSnBr3} and \ce{Cs2SnBr6}. Molecular Dynamics simulations in the $NpT$ ensemble were performed from 100 K to 500 K, and thermodynamic and structural descriptors including enthalpy, specific heat, radial distribution functions, translational order, bond angle distributions, and vibrational spectra were analyzed. Our results show that \ce{CsSnBr3} undergoes a low-temperature orthorhombic-to-cubic phase transition, evidenced by both the evolution of lattice parameters and subtle anomalies in enthalpy and specific heat, although the experimentally observed intermediate tetragonal phase is not captured. In contrast, \ce{Cs2SnBr6} remains cubic and maintains a more rigid octahedral framework across the entire temperature range. Overall, MACE-MP-0 qualitatively reproduces key thermal and structural features of these materials, highlighting its usefulness as a first step for studying new materials. For cases where capturing more subtle phase behavior is required, system-specific fine-tuning with Density Functional Theory data should be considered.

\end{abstract}

\maketitle

\section{Introduction}

In materials chemistry, the so-called halide perovskites have emerged as a key class of materials due to their structural diversity and compositional tunability~\cite{liang2023structural}. These materials share the same crystalline structure as calcium titanate (\ce{CaTiO3})\cite{inamuddin2019green}, where, in standard \ce{ABX3} compounds, the A-site cation occupies a cuboctahedral void formed by corner-sharing \ce{BX6} octahedra. When X is a halide, and A and B are monovalent and divalent cations, respectively, the resulting compound is referred to as a halide perovskite. These materials have attracted great attention in optoelectronics, particularly photovoltaics, with solar cell efficiencies now exceeding 25\%\cite{ansari2018frontiers,dale2023efficiency,li2025tin}. However, despite their remarkable performance, their widespread application is hindered by issues related to chemical and structural stability~\cite{frost2016moving,carignano2017critical,jinnouchi2019phase,wiktor2023quantifying,yi2016entropic,beal2016cesium,raval2022understanding}.

One particularly pressing concern is the environmental toxicity associated with lead (\ce{Pb}), which is commonly used as the B-site cation in high-performance perovskites~\cite{ren2022potential}. Although silicon remains dominant in the solar cell market, any transition toward lead-based perovskites must confront this sustainability challenge. A promising strategy is to substitute lead with less toxic alternatives, such as \ce{Sn^2+}, \ce{Ge^2+}, or other divalent metals~\cite{hoefler2017progress,jena2019halide}. Among these, tin (Sn) has garnered significant interest due to its similar valence and ionic radius, and its favorable semiconducting properties~\cite{jena2019halide,ke2019unleaded}. Nevertheless, tin-based perovskites still face challenges in terms of phase stability~\cite{aktas2022challenges}, degradation~\cite{lanzetta2020stability} and oxidizing process~\cite{dalpian2017changes}. A detailed understanding of their atomic-scale dynamics is thus essential to improving their performance and robustness.

Atomic-scale simulations offer powerful means of studying these materials and their complex structural behavior. In particular, Density Functional Theory (DFT) has become an essential method for accurately modeling structural, thermodynamic, and electronic properties~\cite{kohn1965self,kresse1993ab,giannozzi2020quantum,kuhne2020cp2k}. This approach, though, becomes challenging for halide perovskites, whose functional performance is often linked to anharmonic lattice dynamics, soft phonon modes, and temperature-driven phase transitions~\cite{fransson2023revealing,farigliano2024phase}. Since the presence of soft modes implies a strong coupling between lattice distortions and temperature, these systems can be investigated through static 0~K calculations using polymorphic structures~\cite{zhao2020polymorphous} to explore their potential-energy landscape, even though such models do not fully capture the real behavior of the material. To explicitly include finite-temperature effects and obtain dynamic information about the system, \textit{ab initio} Molecular Dynamics simulations are required~\cite{farigliano2024phase,kaiser2021halide,kaiser2022stability,mosconi2015ab,carignano2015thermal}. These methods, nevertheless, come with a high computational cost that limits their application to large-scale or long-time simulations.

To address these challenges, machine learning (ML) interatomic potentials have emerged as a highly promising solution. These methods offer near-DFT accuracy at a fraction of the cost, provided that robust training datasets and modeling frameworks are available~\cite{behler2007generalized,bartok2010gaussian,thompson2015spectral,schutt2018schnet,deringer2019machine,batzner20223,ko2023recent}. Among them, the Multi-scale Atomic Cluster Expansion (MACE)~\cite{batatia2023foundation,batatia2025design} has recently demonstrated exceptional performance. MACE represents interatomic interactions via high-order, symmetrized atomic clusters and relies on compact message-passing schemes that encode four-body correlations. This architecture allows MACE to achieve DFT-level accuracy with reduced training set sizes and lower computational complexity~\cite{drautz2019atomic}. To extend its transferability, a foundational version of the model, MACE-MP-0, has been trained on a large dataset from the Materials Project~\cite{jain2016computational}. This general-purpose model can predict energies and forces across a wide chemical space.

In this study, we investigate the ability of the MACE-MP-0 foundational model — trained on general-purpose materials data — to describe the structural and thermal behavior of tin-based halide perovskites \ce{CsSnBr3} and \ce{Cs2SnBr6}, promising lead-free candidates for optoelectronic applications. Using Molecular Dynamics simulations in the $NpT$ ensemble over a temperature range of 100 K to 500 K, we evaluate the model’s ability to capture temperature-driven phenomena such as structural distortions, octahedral tilting, and phase transitions~\cite{fransson2023revealing,hirotsu1974structural,stoumpos2013crystal,onoda1990calorimetric,whitfield2016structures,mao2025correlated,liang2025phase}. We analyze structural and thermodynamic descriptors including lattice parameters, enthalpy, specific heat, radial distribution functions, translational order, angular distributions, and vibrational spectra. Our results reveal that \ce{CsSnBr3} undergoes a low-temperature orthorhombic-to-cubic phase transition, although the experimentally observed intermediate tetragonal phase is not captured, while \ce{Cs2SnBr6} remains cubic with a notably rigid octahedral framework. These findings demonstrate that MACE-MP-0, even without task-specific fine-tuning, can qualitatively reproduce key features of halide perovskite behavior, highlighting its usefulness as a first step for studying new materials. For more subtle phase behavior, system-specific fine-tuning with DFT data may be necessary.

This work is organized as follows: in Section~\ref{methods}, we describe the simulation methodology, including structural models, ensemble conditions, and analysis protocols. Section~\ref{results} presents and discusses the results of the molecular dynamics simulations, focusing on the structural response and thermodynamic behavior of both perovskites. Finally, our Conclusions summarizes the main findings and offers perspectives on the use of foundational machine learning models in atomistic simulations of complex materials.

\section{Simulation Details and Analysis Methods}\label{methods}

\begin{figure}[h!]
    \centering
        \includegraphics[height=4cm]{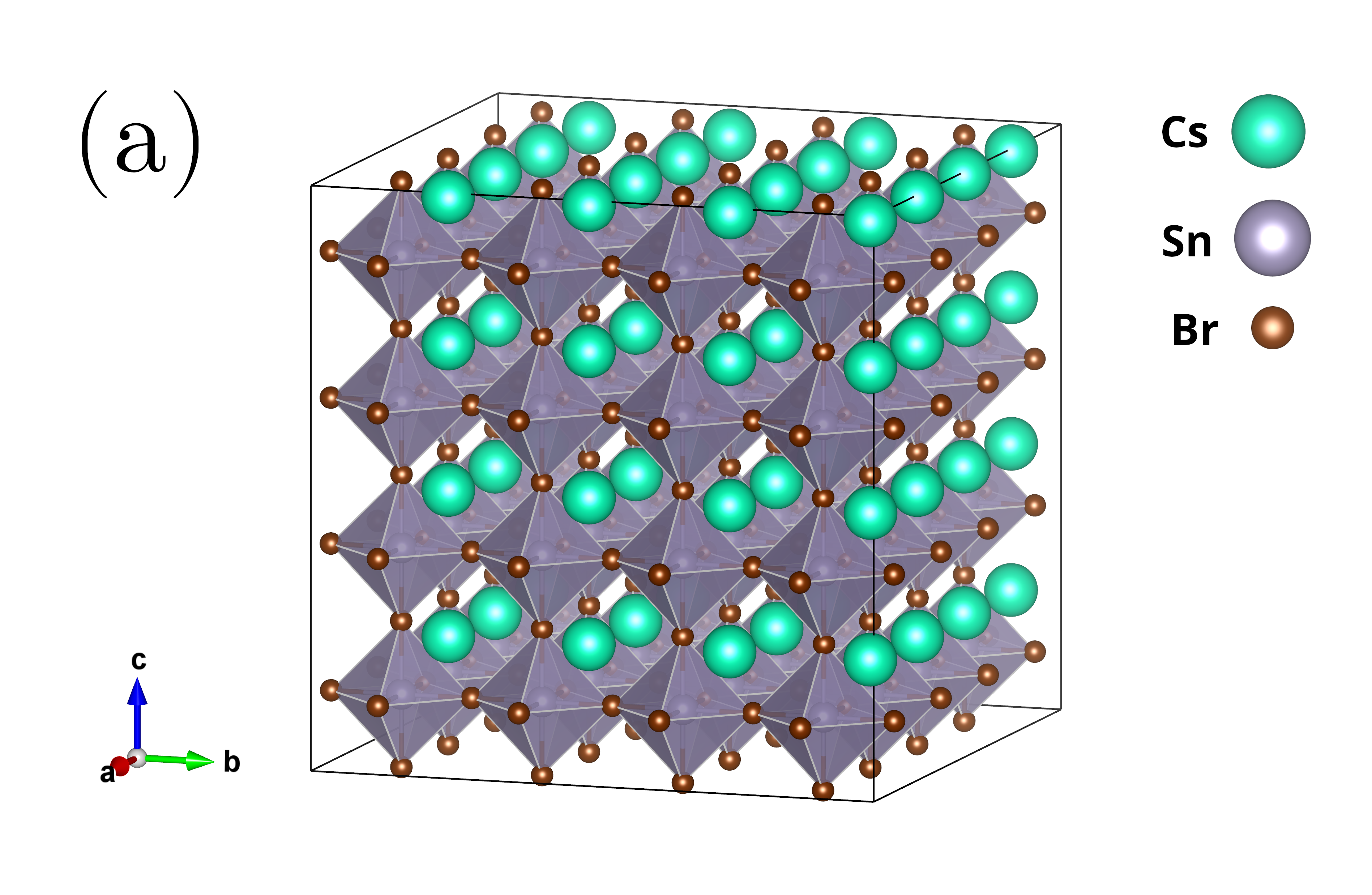}
        \includegraphics[height=4cm]{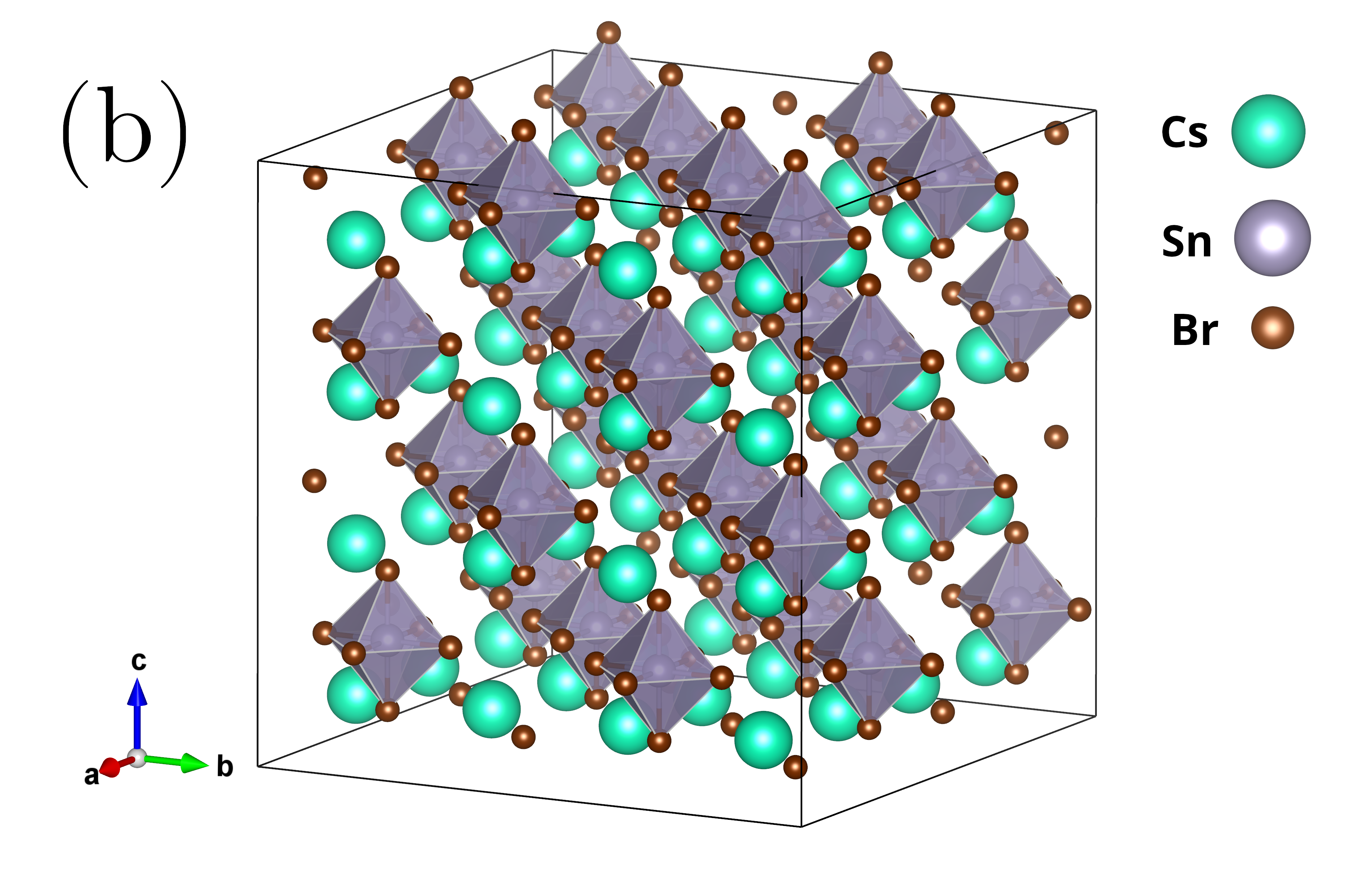}
    \caption{Crystal structures of the simulated supercells: (a) cubic \ce{CsSnBr3} and (b) cubic \ce{Cs2SnBr6}, both with $4 \times 4 \times 4$ replication.}
    \label{fig: system-snap}
\end{figure}

We performed Molecular Dynamics (MD) simulations in the $NpT$ ensemble for two systems: (i) a cubic \ce{CsSnBr3} $4 \times 4 \times 4$ supercell and (ii) a cubic \ce{Cs2SnBr6} $4 \times 4 \times 4$ supercell, as illustrated in Fig.~\ref{fig: system-snap}~(a) and (b), respectively. The simulations were carried out using the LAMMPS package~\cite{lammps}. The interatomic interactions between \ce{Cs}, \ce{Sn}, and \ce{Br} atoms were modeled using the MACE-MP-0 foundational model~\cite{batatia2023foundation}.

We followed a heating protocol from $T = 100$~K to $T = 500$~K, in increments of $\Delta T = 50$~K. The pressure was maintained constant at $p = 1.0$~bar. For each temperature, we ran 2~ns of simulation time, where the first 1~ns was used for equilibration and the remaining 1~ns for data production. The timestep was set to $\delta t = 0.001$~ps. Temperature and pressure were controlled using the Nosé–Hoover thermostat and barostat, with coupling constants $Q_T = 0.1$~ps and $Q_p = 0.5$~ps, respectively. As the temperature increased along the isobaric heating pathway, the final configuration from each run was used as the initial configuration for the subsequent temperature step. Thermodynamic, dynamic, and structural observables were monitored throughout the simulations.

Thermodynamic quantities such as the system enthalpy $H$,
\begin{equation}
H = U + pV,
\label{eq: enthalpy}
\end{equation}
where $U$ is the internal energy, $p$ is the pressure, and $V$ is the system volume, were monitored. To investigate potential phase transitions along the isobar, we computed the specific heat at constant pressure $c_p$ via
\begin{equation}
c_p = \frac{1}{N} \left( \frac{\partial H}{\partial T} \right)_p,
\label{eq: cp}
\end{equation}
where $N$ is the total number of atoms in the system. These derivatives were evaluated numerically, and as a consistency check, we verified that the locations of maxima matched those obtained via statistical fluctuations~\cite{allen2017}.

Structural properties such as supercell angles and lattice parameters were evaluated during the production phase. We also computed the radial distribution function (RDF) to characterize local structural order, defined as
\begin{equation}
g(r) = \frac{1}{\rho} \left< \frac{1}{N} \sum_{i=1}^{N} \sum_{j=i+1}^{N} \delta(\mathbf{r} - \mathbf{r}_{ij}) \right>,
\label{eq_gr}
\end{equation}
where $\rho$ is the average number density and $\mathbf{r}_{ij}$ is the distance between particles $i$ and $j$. In addition, we calculated the translational order parameter~\cite{Errington2001}, given by
\begin{equation}
\tau = \frac{1}{s_c} \int_{0}^{s_c} |g(r) - 1| \, dr,
\end{equation}
where $s_c$ is a cutoff radius beyond which $g(r)$ oscillations become negligible.

In order to further assess octahedral distortions, we analyzed the internal structure of the \ce{SnBr6} units by computing \ce{Br}-\ce{Sn}-\ce{Br} bond angles. We identified all pairs of bromine atoms bonded to the same tin atom within a cutoff distance and calculated the angle $\theta$ using the cosine law:
\begin{equation}
\theta = \cos^{-1} \left( \frac{\vec{r}_{i} \cdot \vec{r}_{j}}{|\vec{r}_{i}| \, |\vec{r}_{j}|} \right),
\end{equation}
where $\vec{r}_{i}$ and $\vec{r}_{j}$ are vectors from the central \ce{Sn} atom to two distinct neighboring \ce{Br} atoms. In addition to the average angle, we analyzed the distribution of \ce{Br}-\ce{Sn}-\ce{Br} angles and extracted its full width at half maximum (FWHM). This metric serves as a quantitative measure of angular dispersion, reflecting the extent of thermal or structural distortions within the octahedra.

The vibrational spectrum was obtained from the molecular dynamics trajectories using the velocity autocorrelation function (VACF) formalism. The normalized VACF is defined as

\begin{equation}
C_v(t) = \frac{\langle \mathbf{v}(0) \cdot \mathbf{v}(t) \rangle}{\langle \mathbf{v}(0) \cdot \mathbf{v}(0) \rangle},
\end{equation}
where $\mathbf{v}(t)$ represents the instantaneous atomic velocities at time $t$, and the brackets $\langle \cdots \rangle$ denote an ensemble or time average over all atoms and time origins.

The vibrational density of states (VDOS) was obtained from the Fourier transform of the VACF as

\begin{equation}
g(\omega) = \int_{-\infty}^{+\infty} C_v(t)\, e^{-i\omega t} \, dt,
\end{equation}
which provides the distribution of vibrational frequencies $\omega$ sampled during the dynamics. To obtain a quantity proportional to the infrared (IR) absorption spectrum, the Fourier-transformed data were further multiplied by $\omega^2$, since the IR intensity scales with the square of the vibrational frequency:

\begin{equation}
I(\omega) \propto \omega^2 g(\omega).
\end{equation}

This approach inherently includes anharmonic effects and finite-temperature contributions, as it is based on molecular dynamics trajectories rather than harmonic approximations. It is grounded on the well-established relation between the Fourier transform of the velocity autocorrelation function and the vibrational density of states in harmonic systems, as discussed by Allen and Tildesley~\cite{allen2017}. In practice, the VACF was computed from the atomic velocity data along the trajectory, averaged over all atoms and time origins, and its discrete Fourier transform was evaluated to obtain $g(\omega)$ and subsequently $I(\omega)$.

\section{Results and Discussion}\label{results}

\begin{figure}[h!]
    \centering
        \includegraphics[height=4.5cm]{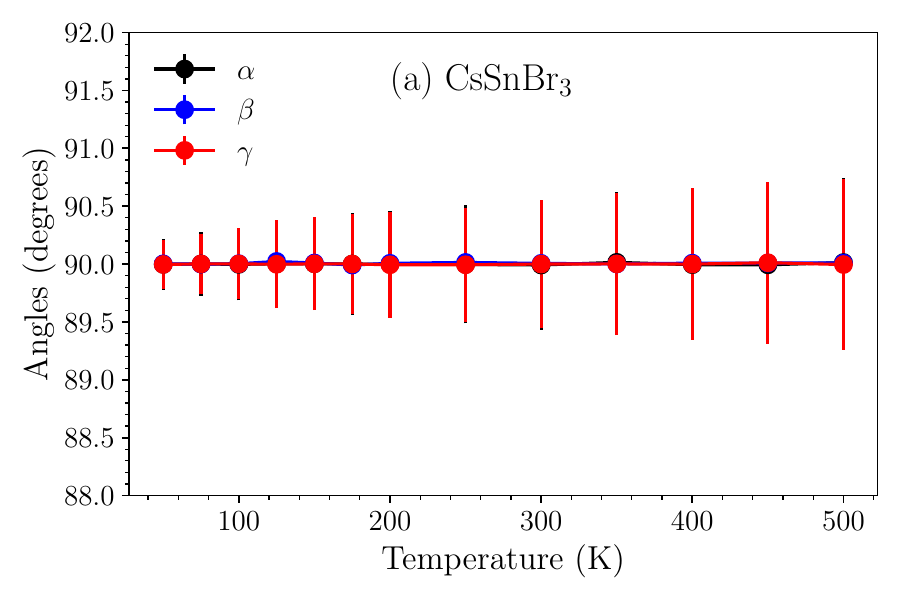}
        \includegraphics[height=4.5cm]{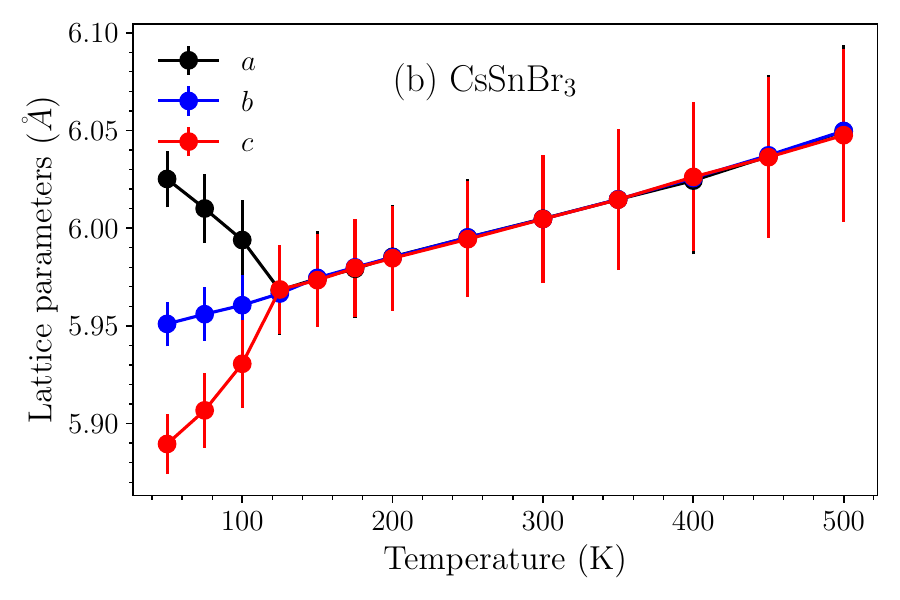}
        \includegraphics[height=4.5cm]{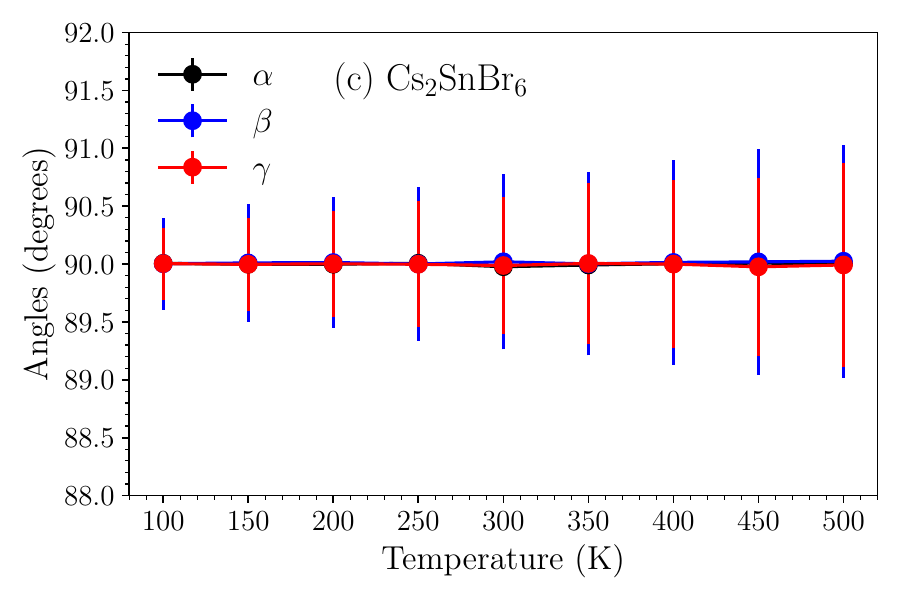}
        \includegraphics[height=4.5cm]{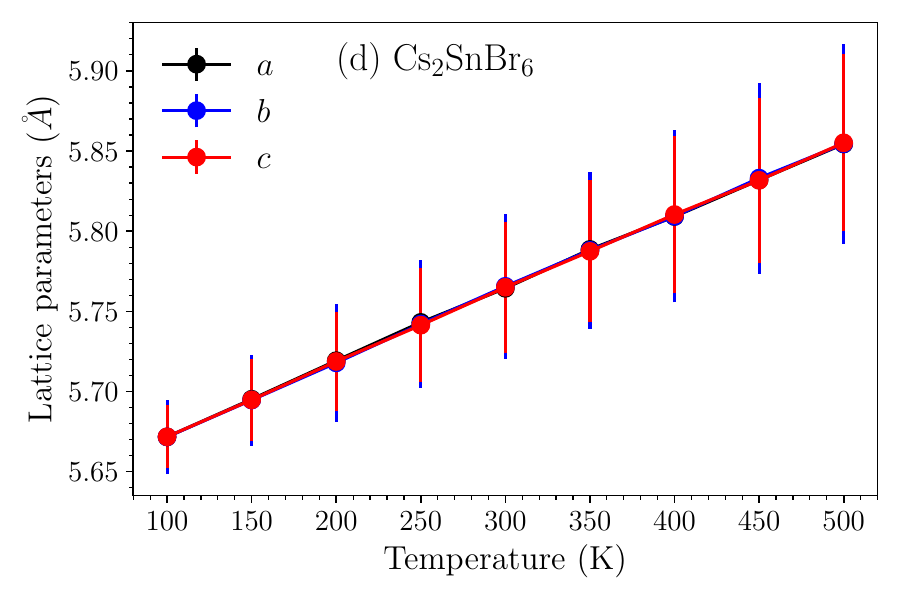}
    \caption{Average values obtained during the production stage ($T = 100$–$500$ K): (a) cell angles $\alpha$, $\beta$, and $\gamma$ for \ce{CsSnBr3}; (b) lattice parameters $a$, $b$, and $c$ normalized per octahedral unit for \ce{CsSnBr3}; (c) cell angles $\alpha$, $\beta$, and $\gamma$ for \ce{Cs2SnBr6}; and (d) lattice parameters $a$, $b$, and $c$ normalized per octahedral unit for \ce{Cs2SnBr6}.}
    \label{fig: angles_lattice_allT}
\end{figure}

In this work we aim to investigate the predictions of MACE-MP-0 for Sn-based halide Perovskites without incurring on fine-tuning it to home computed Density Functional Theory (DFT) dataset. The heating simulations performed from $T = 100$ K to $T = 500$ K ($\Delta T = 50$ K) allowed us to analyze the structural evolution of \ce{CsSnBr3} and \ce{Cs2SnBr6} as a function of temperature. Fig.~\ref{fig: angles_lattice_allT} shows, respectively, the angles (a) and lattice parameters normalized per octahedral unit (b) for \ce{CsSnBr3}, while the equivalent data for \ce{Cs2SnBr6} are shown in Fig.~\ref{fig: angles_lattice_allT} (c) and (d).

At temperatures above 150 K, both systems preserved a cubic symmetry on average, as indicated by the equality of the cell angles ($\alpha = \beta = \gamma$) and lattice parameters ($a = b = c$). However, from $T = 50$ K up to $T = 125$ K, a clear distortion is observed for \ce{CsSnBr3}, with $a \neq b \neq c$ while $\alpha = \beta = \gamma = 90^\circ$, indicating an orthorhombic arrangement. In contrast, \ce{Cs2SnBr6} remains cubic across this temperature range. This behavior suggests that \ce{CsSnBr3} undergoes a structural phase transition at low temperatures, consistent with experimental reports. X-ray diffraction measurements~\cite{mori1986x} identified a monoclinic-to-cubic transition in this temperature range, while neutron diffraction studies~\cite{fabini2016dynamic} refined the sequence of transitions as orthorhombic at 100 K, tetragonal at 270 K, and cubic at 300 K.

Table \ref{tab: cssnbr3} presents the unit cell lattice parameters of \ce{CsSnBr3} obtained from simulations at different temperatures, alongside experimental data extracted from the literature~\cite{fabini2016dynamic}. At 50 K, the theoretical values reproduce well the orthorhombic \textit{Pnma} structure, with the characteristic elongation of the $b$ axis and contraction of the $a$ and $c$ axes relative to the pseudo-cubic cell. As the temperature increases, the simulated structure undergoes a direct transition to the cubic phase, as reflected in the convergence of the lattice parameters. In contrast, the experimental data shows that at 270 K this material exhibits a tetragonal \textit{P4/mbm} phase. Overall, despite the absence of this phase in the simulations, the temperature-dependent evolution of the lattice parameters agrees well with the experimental trends.

For \ce{Cs2SnBr6}, the calculated lattice parameters at different temperatures are summarized in Table \ref{tab: cs2snbr6}. Throughout the investigated temperature range, the structure remains cubic with space group \textit{Fm-3m}, exhibiting an isotropic thermal expansion of the lattice parameter and unit cell volume. To the best of our knowledge, experimental data for this compound at the considered temperatures are not available in the literature, preventing a direct comparison. Nevertheless, the theoretical results provide a consistent structural reference for \ce{Cs2SnBr6}.

\begin{table}[h!]

\centering
\caption{Comparison of theoretical lattice parameters and volume obtained from our simulations with experimental values for \ce{CsSnBr3}. Theoretical values correspond to unit cell parameters. Experimental data (*) are taken from Ref.~\cite{fabini2016dynamic}.}
\begin{tabular}{lcccccc}
\hline
                & \multicolumn{3}{c}{Theoretical}                 & \multicolumn{3}{c}{Experimental*}                 \\ \hline
temperature (K) & 50            & 125            & 300            & 100           & 270             & 300            \\
space group     & \textit{Pnma} & \textit{Pm-3m} & \textit{Pm-3m} & \textit{Pnma} & \textit{P4/mbm} & \textit{Pm-3m} \\
a (\AA)           & 8.32        & 5.96         & 6.00         & 8.1965        & 8.1789          & 5.8043         \\
b (\AA)           & 11.90       & 5.96         & 6.00         & 11.5830       & 8.1798          & 5.8043         \\
c (\AA)           & 8.52       & 5.96         & 6.00         & 8.0243        & 5.8193          & 5.8043         \\
volume (\AA$^3$)     & 844.68        & 212.52         & 216.48         & 761.82        & 389.29          & 195.54        \\ \hline
\end{tabular}
\label{tab: cssnbr3}
\end{table}

\begin{table}[h!]
\small
\centering
\caption{Theoretical lattice parameters and volume obtained from our simulations for \ce{Cs2SnBr6}.These values correspond to unit cell parameters.}
\begin{tabular}{lccc}
\hline
                & \multicolumn{3}{c}{Theoretical}                  \\ \hline
temperature (K) & 100            & 200            & 300            \\
space group     & \textit{Fm-3m} & \textit{Fm-3m} & \textit{Fm-3m} \\
a (\AA)           & 11.34        & 11.43        & 11.53        \\
b (\AA)           & 11.34        & 11.43        & 11.53        \\
c (\AA)           & 11.34        & 11.43        & 11.53        \\
volume (\AA$^3$)     & 1459.52        & 1495.95        & 1533.08        \\ \hline
\end{tabular}
\label{tab: cs2snbr6}
\end{table}

\begin{figure}[h]
    \centering
        \includegraphics[height=5.2cm]{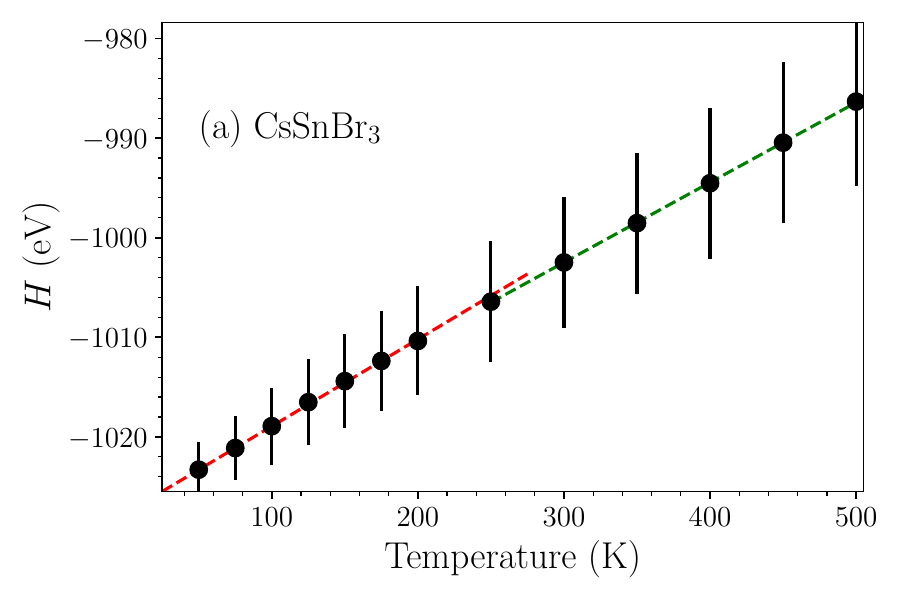}
        \includegraphics[height=5.2cm]{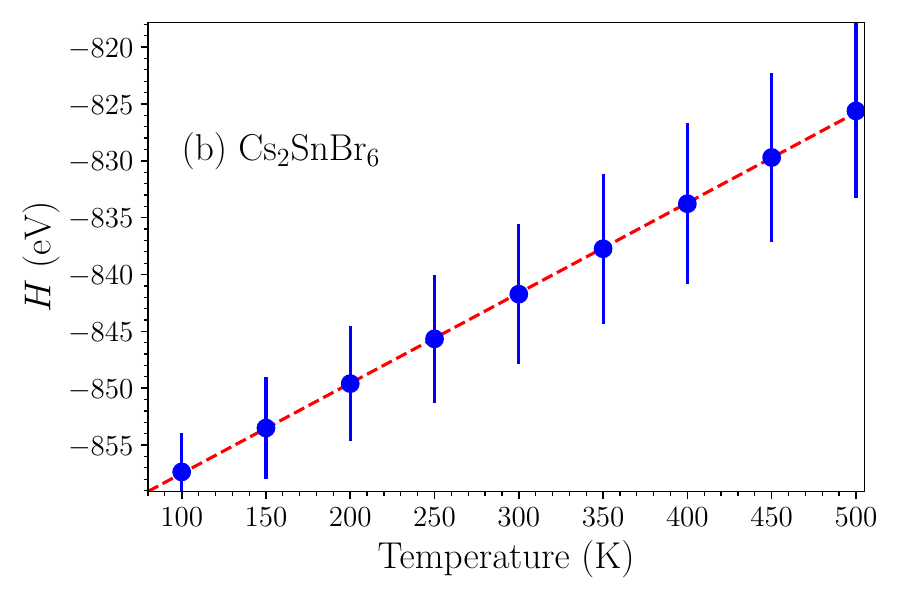}
        \includegraphics[height=5.2cm]{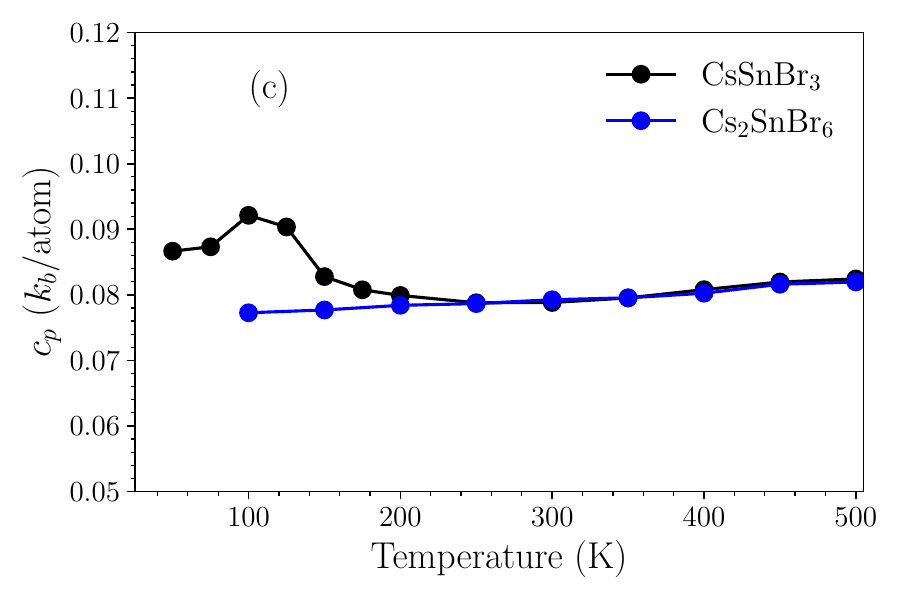}
        \caption{(a) \ce{CsSnBr3} enthalpy $H$ as function of the system's temperature $T$, (b) \ce{Cs2SnBr6} enthalpy $H$ as function of the system's temperature $T$ and (c) the specific heat at constant pressure $c_p$ for both systems as function of the temperature $T$. The lines are for guiding the eyes.}
    \label{fig: enthalpy_cp}
\end{figure}

To probe the phase transition, we computed the enthalpy $H$ as a function of temperature for both systems (Fig.~\ref{fig: enthalpy_cp} (a) and (b)). As expected, $H$ increases with temperature due to thermal fluctuations, with larger error bars at higher $T$. For \ce{CsSnBr3}, a subtle deviation from linearity emerges between 50 K and 300 K (highlighted by the dashed red and green lines in Fig.~\ref{fig: enthalpy_cp} (a)), suggestive of a second-order phase transition. Analysis of lattice parameters and cell angles indicates that \ce{CsSnBr3} evolves from an orthorhombic arrangement at 100 K ($a \neq b \neq c$; $\alpha = \beta = \gamma = 90^\circ$) toward cubic symmetry at higher temperatures, consistent with the enthalpy trend. To corroborate this transition, we evaluated the specific heat at constant pressure, $c_p$ (Equation~\ref{eq: cp}), for both systems (Fig.~\ref{fig: enthalpy_cp} (c)). In contrast to the cubic \ce{Cs2SnBr6}, which displays an essentially constant $c_p$, \ce{CsSnBr3} exhibits a clear peak near 100 K followed by stabilization around 250 K. This feature signals the orthorhombic-to-cubic transition and aligns with both the structural evolution and experimental reports of low-temperature transitions in \ce{CsSnBr3}~\cite{mori1986x,fabini2016dynamic}. Overall, these results demonstrate that the MACE-MP-0 model, even without fine-tuning to system-specific DFT data, qualitatively captures the low-temperature orthorhombic-to-cubic transition of \ce{CsSnBr3}. The model, however, does not reproduce the experimentally observed intermediate orthorhombic-to-tetragonal transition near 270 K~\cite{fabini2016dynamic}, indicating that system-specific fine-tuning could be beneficial for capturing more subtle phase behavior. Nevertheless, these findings highlight the value of MACE-MP-0 as a general-purpose tool for exploring the structural and thermal properties of halide perovskites.

\begin{figure}[h]
    \centering
        \includegraphics[height=5cm]{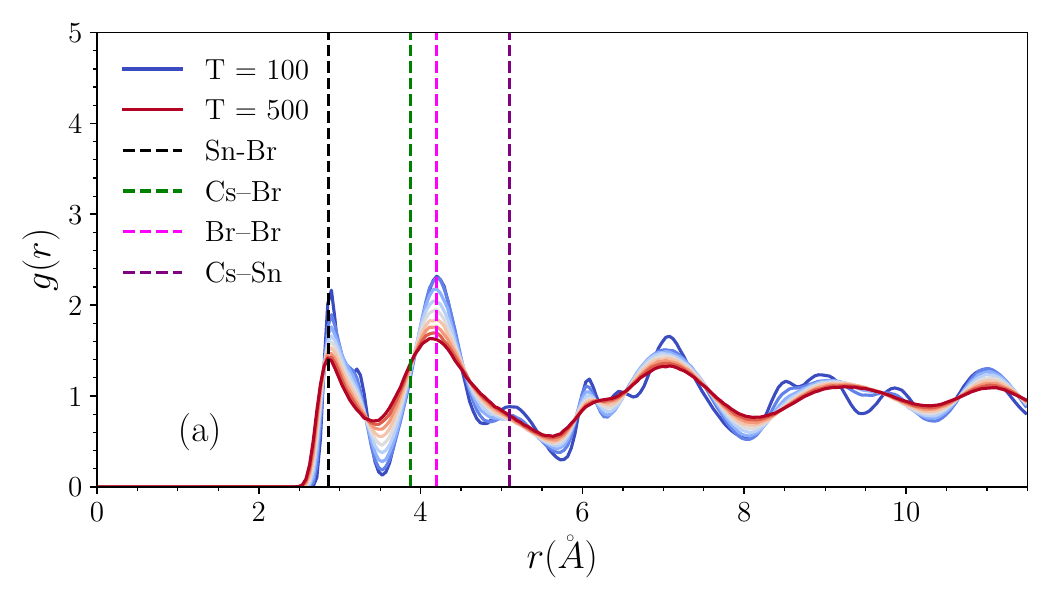}
        \includegraphics[height=5cm]{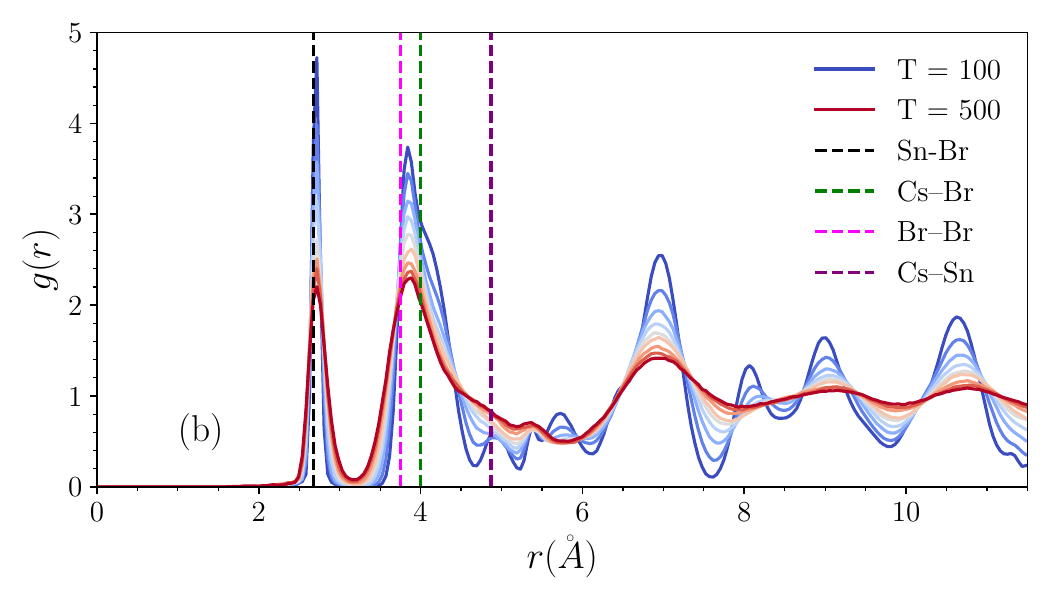}
        \caption{Radial distribution function $g(r)$ for (a) \ce{CsSnBr3} and (b) \ce{Cs2SnBr6} computed along the isobar from 100~K to 500~K. The curves highlight how thermal fluctuations progressively broaden and lower the peaks, especially beyond the first coordination shell. Vertical dashed lines indicate characteristic interatomic distances at 0~K.}
    \label{fig: rdfall}
\end{figure}

Structural properties were computed along the isobars. In Fig.~\ref{fig: rdfall} (a) and (b), we show the radial distribution function $g(r)$ for \ce{CsSnBr3} and \ce{Cs2SnBr6}, respectively, for temperatures ranging from 100~K to 500~K. As the temperature increases, thermal fluctuations significantly affect both structures. This is evidenced by the broadening and lowering of the $g(r)$ peaks, particularly beyond the first coordination shell. Such behavior reflects a loss of medium- and long-range order due to increased atomic vibrations and dynamic disorder. For \ce{CsSnBr3}, the first peak around $r \approx 2.8$~\AA{}, corresponding to the Sn--Br bond, remains sharp and well-defined even at 500~K, indicating persistent short-range order. However, the higher-order peaks (e.g., around 4--6~\AA) become increasingly diffuse with temperature. In contrast, \ce{Cs2SnBr6} exhibits a more intense and narrower first peak at the same Sn--Br distance, with values exceeding 4.5 at 100~K. This suggests a more rigid local octahedral environment, which remains preserved even at high temperatures. Nonetheless, both systems show a clear reduction in the amplitude and definition of peaks at longer distances, indicating a gradual loss of translational symmetry.

\begin{figure}
    \centering
    \includegraphics[height=5cm]{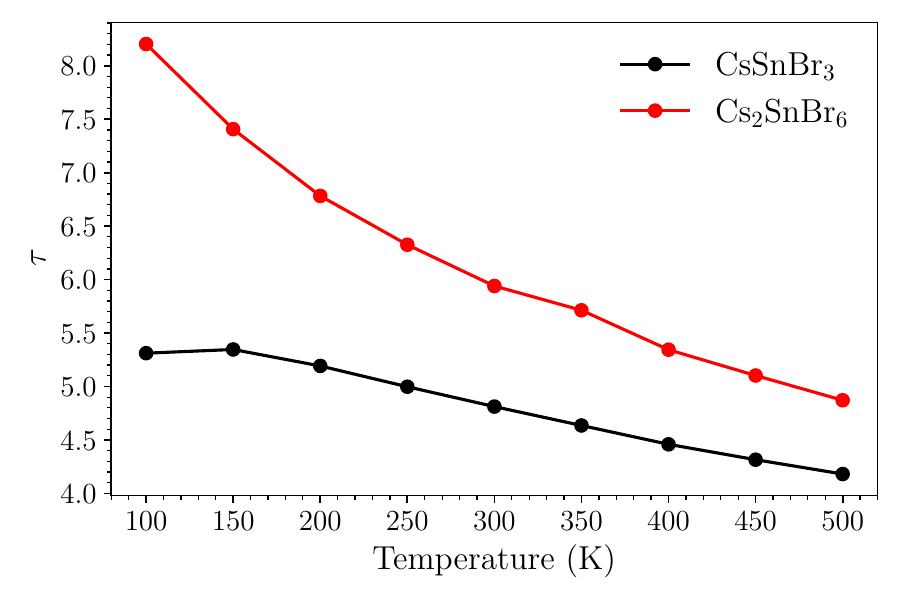}
    \caption{Translational order parameter $\tau$ as a function of temperature for \ce{CsSnBr3} (black) and \ce{Cs2SnBr6} (red). Both systems exhibit a monotonic decrease in $\tau$, indicating loss of positional order. \ce{Cs2SnBr6} consistently shows higher values of $\tau$, suggesting greater structural coherence.}
    \label{fig:tau}
\end{figure}

In order to further quantify the structural order, we computed the translational order parameter $\tau$ (Fig.~\ref{fig:tau}). In both materials, $\tau$ decreases monotonically with temperature, confirming the progressive loss of positional correlations. Interestingly, \ce{Cs2SnBr6} exhibits consistently higher $\tau$ values than \ce{CsSnBr3}, suggesting that it retains a greater degree of structural coherence throughout the temperature range. This trend agrees with the $g(r)$ analysis and reflects the stronger bonding and more rigid framework of the \ce{SnBr6} octahedra. Overall, the combined analysis of $g(r)$ and $\tau$ indicates that both perovskites maintain short-range structural motifs up to 500K, while gradually losing medium- and long-range order. These results highlight the thermal resilience of the octahedral units, particularly in \ce{Cs2SnBr6}, and provide insight into the microscopic mechanisms underlying structural stability in these materials.

\begin{figure}[h!]
    \centering
    \includegraphics[height=5cm]{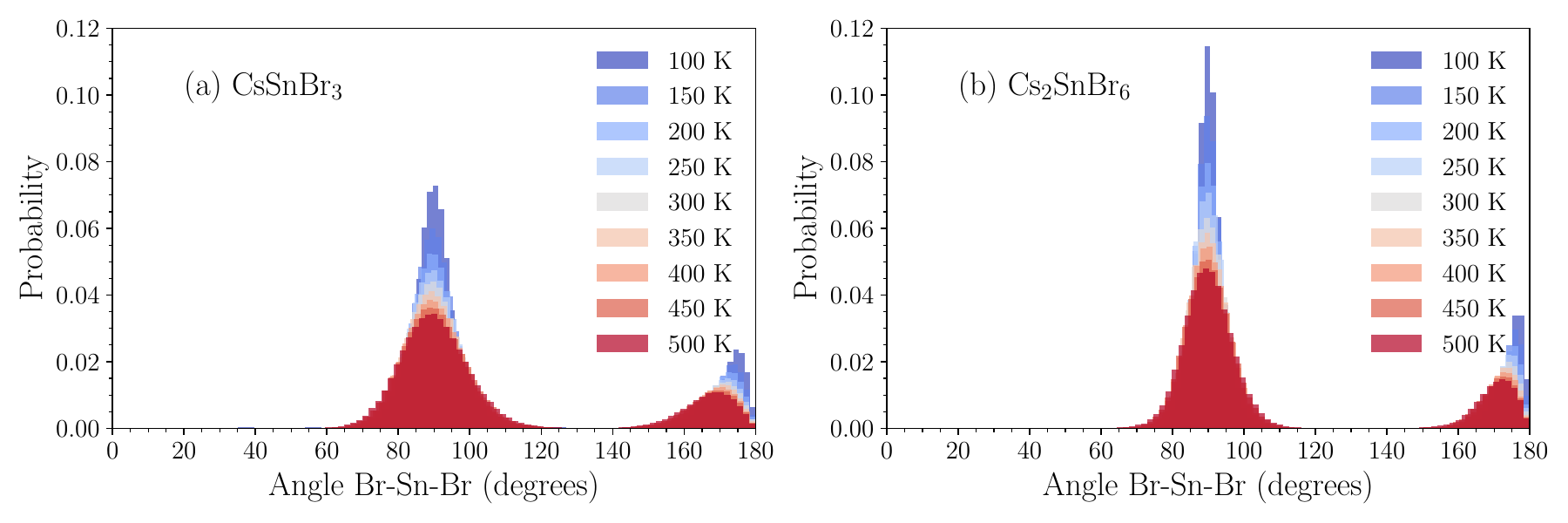}
    \caption{Probability distribution of the Br--Sn--Br bond angle for (a) \ce{CsSnBr3} and (b) \ce{Cs2SnBr6}, at temperatures ranging from 100~K to 500~K. The peaks near 90° and 180° are associated with intra- and inter-octahedral angles, respectively. \ce{Cs2SnBr6} shows narrower and more stable distributions, indicating greater octahedral rigidity.}
    \label{fig:angle_distribution}
\end{figure}

\begin{figure}[h!]
    \centering
        \includegraphics[height=5cm]{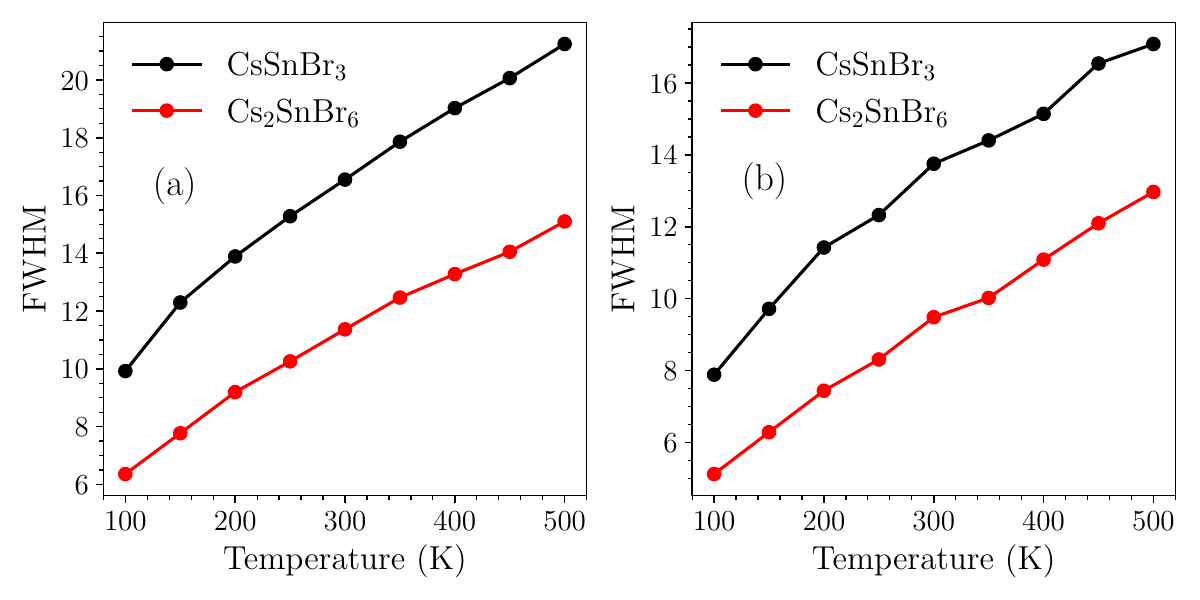}
        \includegraphics[height=5cm]{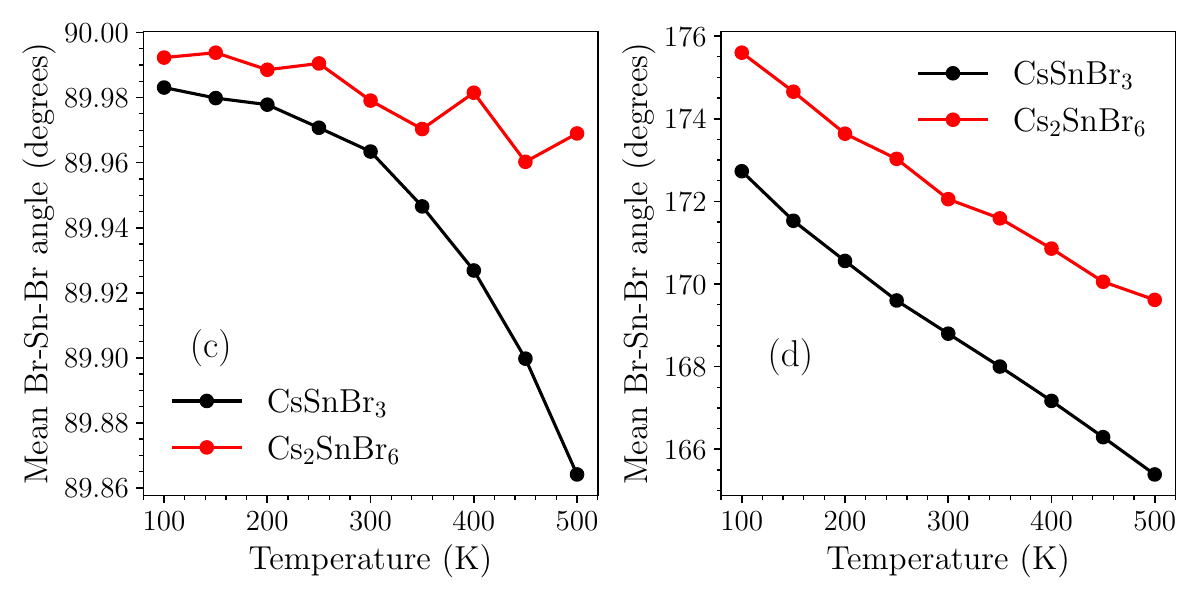}
        \caption{Temperature dependence of angular properties for \ce{CsSnBr3} (black) and \ce{Cs2SnBr6} (red): 
    (a) FWHM of the Br--Sn--Br peak near 90°, (b) FWHM near 180°, 
    (c) mean Br--Sn--Br angle near 90°, and (d) mean Br--Sn--Br angle near 180°. The octahedra in \ce{Cs2SnBr6} display systematically lower angular distortions and reduced thermal sensitivity compared to \ce{CsSnBr3}.}
    \label{fig:angle_metrics}
\end{figure}

To further probe the local structure and octahedral stability, we analyzed the distribution of Br–Sn–Br bond angles for both compounds (Fig.~\ref{fig:angle_distribution}). For \ce{CsSnBr3}, the angular distributions broaden considerably with temperature, especially around 90°, indicating increasing octahedral tilts and dynamic distortions. In contrast, \ce{Cs2SnBr6} maintains sharper distributions centered near 90° and 180°, even at 500K, suggesting a more rigid and symmetric octahedral framework. Fig.~\ref{fig:angle_metrics} quantifies this behavior by showing the full width at half maximum (FWHM) and the average values of the angular peaks. Both the FWHM and the deviation from ideal angles increase more significantly for \ce{CsSnBr3}, reinforcing that its structure is more susceptible to thermal distortions. The higher angular stability in \ce{Cs2SnBr6} complements the previous analyses of $g(r)$ and $\tau$, consolidating the view that the octahedral units in this phase are thermally more resilient.

\begin{figure}[h]
    \centering
    \includegraphics[height=5cm]{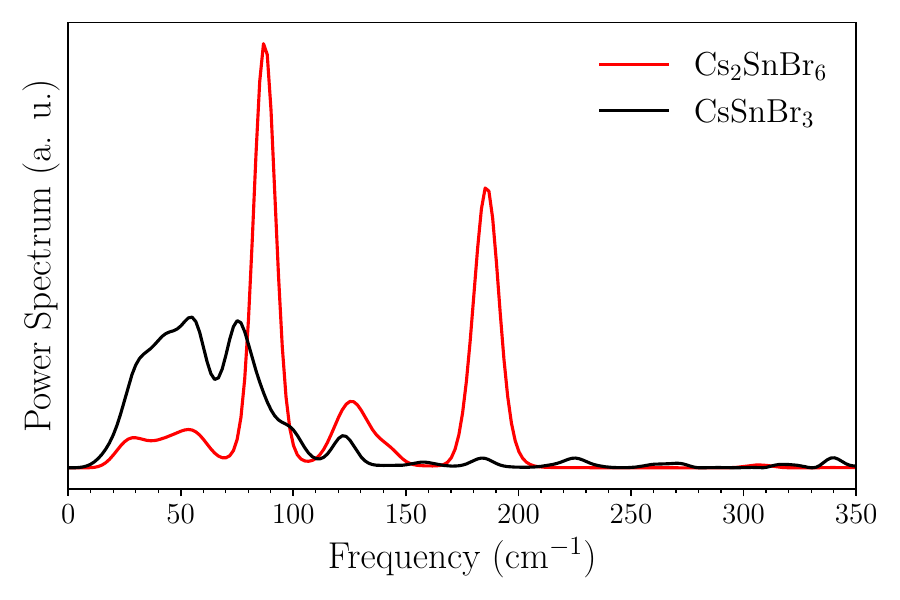}
    \caption{Calculated vibrational spectra of the full structures of \ce{Cs2SnBr6} (red) and \ce{CsSnBr3} (black). Sharper and higher-frequency peaks in \ce{Cs2SnBr6} indicate a dynamically stiffer lattice, consistent with its more rigid octahedral framework.} 
    \label{fig: spectrum}
\end{figure}

Beyond the static structural descriptors, additional insight into the lattice rigidity can be obtained from the vibrational spectra (Fig.~\ref{fig: spectrum}). These spectra represent the collective vibrational response of the entire lattice, combining both local and extended modes. Despite this global character, marked differences emerge between the two compounds. \ce{Cs2SnBr6} exhibits sharper and more intense peaks at higher frequencies, while \ce{CsSnBr3} shows broader and more pronounced low-frequency features. Such spectral contrasts indicate that the overall lattice dynamics of \ce{Cs2SnBr6} are stiffer, with more localized vibrational modes and reduced anharmonicity, whereas \ce{CsSnBr3} exhibits softer, more collective low-frequency motions associated with octahedral tilting and framework flexibility. These findings are consistent with the later results confirming the higher rigidity of the octahedral units in \ce{Cs2SnBr6}.

\section{Conclusion}\label{conclusions} 

In this study, we have demonstrated that the MACE-MP-0 model, even without system-specific fine-tuning to our DFT dataset, is capable of qualitatively capturing key thermal and structural behaviors of the Sn-based halide perovskites \ce{CsSnBr3} and \ce{Cs2SnBr6}. Heating simulations from 100 K to 500 K reveal that both materials maintain cubic symmetry on average at elevated temperatures, while \ce{CsSnBr3} exhibits a low-temperature distortion consistent with an orthorhombic-to-cubic phase transition reported experimentally. However, the model does not reproduce the experimentally observed intermediate tetragonal phase of \ce{CsSnBr3}, indicating that fine-tuning with system-specific DFT data could improve the description of subtle phase behavior.

Thermodynamic analysis, including enthalpy and specific heat calculations, supports the presence of a gradual second-order phase transition in \ce{CsSnBr3} around 100–300 K, which is absent in \ce{Cs2SnBr6}. Structural metrics such as radial distribution functions, translational order parameters, and bond angle distributions further illustrate the progressive loss of medium- and long-range order with increasing temperature, while preserving short-range octahedral motifs. Notably, \ce{Cs2SnBr6} exhibits higher structural coherence and angular stability of the \ce{SnBr6} octahedra, with sharper radial distribution peaks, higher translational order, and narrower bond angle distributions compared to \ce{CsSnBr3}, confirming its enhanced thermal resilience.

Vibrational spectra reveal that \ce{Cs2SnBr6} displays sharper and higher-frequency peaks, whereas \ce{CsSnBr3} shows broader low-frequency features associated with softer collective motions, corroborating the stiffer lattice and stronger bonding in \ce{Cs2SnBr6}. Taken together, these results demonstrate that MACE-MP-0 captures both static and dynamic aspects of lattice stability in halide perovskites. Overall, the model proves useful as a first step for exploring new materials, while system-specific fine-tuning may be considered when detailed phase behavior or subtle dynamical effects need to be captured.

\section*{Author contributions}
{\bf{Thiago Puccinelli}}: Methodology, Software, Data Curation Validation, Formal analysis, Investigation, Writing - Original Draft. {\bf{Lucas Martin Farigliano}}:  Methodology, Software, Data Curation Validation, Formal analysis, Investigation, Writing - Original Draft. {\bf{Gustavo Martini Dalpian}}: Conceptualization, Methodology, Resources, Writing - Review \& Editing, Supervision, Project administration, Funding acquisition.

\section*{Conflicts of interest}

There are no conflicts to declare

\section*{Data availability}

The MACE-MP-0 model, the initial structures used in this study, and the LAMMPS simulation scripts are publicly available on the corresponding GitHub repository: \url{https://github.com/thiagopuccinelli/mace-mp-o_Sn-based-Halide-Perovskites}. All data required to reproduce the results reported in this work can be accessed there.

\section*{Acknowledgements}
This work has been performed with funding from FAPESP grants 2023/09820-2, 2024/01461-6 and 2024/21550-3. We also would like to acknowledge the Materials Informatics INCT (National Institute of Science and Technology), CNPq, Brazil. This work used resources of the "Centro Nacional de Processamento de Alto Desempenho em S\~{a}o Paulo (CENAPAD-SP)" and Santos Dumont supercomputer at LNCC, Brazil. Portions of the manuscript text were refined with the assistance of OpenAI's ChatGPT, which was used to improve grammar and clarity.

\bibliographystyle{apsrev4-2}
\bibliography{ref}

\begin{thebibliography}{52}%
\makeatletter
\providecommand \@ifxundefined [1]{%
 \@ifx{#1\undefined}
}%
\providecommand \@ifnum [1]{%
 \ifnum #1\expandafter \@firstoftwo
 \else \expandafter \@secondoftwo
 \fi
}%
\providecommand \@ifx [1]{%
 \ifx #1\expandafter \@firstoftwo
 \else \expandafter \@secondoftwo
 \fi
}%
\providecommand \natexlab [1]{#1}%
\providecommand \enquote  [1]{``#1''}%
\providecommand \bibnamefont  [1]{#1}%
\providecommand \bibfnamefont [1]{#1}%
\providecommand \citenamefont [1]{#1}%
\providecommand \href@noop [0]{\@secondoftwo}%
\providecommand \href [0]{\begingroup \@sanitize@url \@href}%
\providecommand \@href[1]{\@@startlink{#1}\@@href}%
\providecommand \@@href[1]{\endgroup#1\@@endlink}%
\providecommand \@sanitize@url [0]{\catcode `\\12\catcode `\$12\catcode `\&12\catcode `\#12\catcode `\^12\catcode `\_12\catcode `\%12\relax}%
\providecommand \@@startlink[1]{}%
\providecommand \@@endlink[0]{}%
\providecommand \url  [0]{\begingroup\@sanitize@url \@url }%
\providecommand \@url [1]{\endgroup\@href {#1}{\urlprefix }}%
\providecommand \urlprefix  [0]{URL }%
\providecommand \Eprint [0]{\href }%
\providecommand \doibase [0]{https://doi.org/}%
\providecommand \selectlanguage [0]{\@gobble}%
\providecommand \bibinfo  [0]{\@secondoftwo}%
\providecommand \bibfield  [0]{\@secondoftwo}%
\providecommand \translation [1]{[#1]}%
\providecommand \BibitemOpen [0]{}%
\providecommand \bibitemStop [0]{}%
\providecommand \bibitemNoStop [0]{.\EOS\space}%
\providecommand \EOS [0]{\spacefactor3000\relax}%
\providecommand \BibitemShut  [1]{\csname bibitem#1\endcsname}%
\let\auto@bib@innerbib\@empty
\bibitem [{\citenamefont {Liang}\ \emph {et~al.}(2023)\citenamefont {Liang}, \citenamefont {Klarbring}, \citenamefont {Baldwin}, \citenamefont {Li}, \citenamefont {Cs{\'a}nyi},\ and\ \citenamefont {Walsh}}]{liang2023structural}%
  \BibitemOpen
  \bibfield  {author} {\bibinfo {author} {\bibfnamefont {X.}~\bibnamefont {Liang}}, \bibinfo {author} {\bibfnamefont {J.}~\bibnamefont {Klarbring}}, \bibinfo {author} {\bibfnamefont {W.~J.}\ \bibnamefont {Baldwin}}, \bibinfo {author} {\bibfnamefont {Z.}~\bibnamefont {Li}}, \bibinfo {author} {\bibfnamefont {G.}~\bibnamefont {Cs{\'a}nyi}},\ and\ \bibinfo {author} {\bibfnamefont {A.}~\bibnamefont {Walsh}},\ }\href@noop {} {\bibfield  {journal} {\bibinfo  {journal} {The Journal of Physical Chemistry C}\ }\textbf {\bibinfo {volume} {127}},\ \bibinfo {pages} {19141} (\bibinfo {year} {2023})}\BibitemShut {NoStop}%
\bibitem [{\citenamefont {Inamuddin}\ \emph {et~al.}(2019)\citenamefont {Inamuddin}, \citenamefont {Asiri},\ and\ \citenamefont {Suvardhan}}]{inamuddin2019green}%
  \BibitemOpen
  \bibfield  {author} {\bibinfo {author} {\bibfnamefont {A.}~\bibnamefont {Inamuddin}}, \bibinfo {author} {\bibfnamefont {A.}~\bibnamefont {Asiri}},\ and\ \bibinfo {author} {\bibfnamefont {K.}~\bibnamefont {Suvardhan}},\ }\href@noop {} {\emph {\bibinfo {title} {Green sustainable process for chemical and environmental engineering and science}}}\ (\bibinfo  {publisher} {Elsevier},\ \bibinfo {year} {2019})\BibitemShut {NoStop}%
\bibitem [{\citenamefont {Ansari}\ \emph {et~al.}(2018)\citenamefont {Ansari}, \citenamefont {Qurashi},\ and\ \citenamefont {Nazeeruddin}}]{ansari2018frontiers}%
  \BibitemOpen
  \bibfield  {author} {\bibinfo {author} {\bibfnamefont {M.~I.~H.}\ \bibnamefont {Ansari}}, \bibinfo {author} {\bibfnamefont {A.}~\bibnamefont {Qurashi}},\ and\ \bibinfo {author} {\bibfnamefont {M.~K.}\ \bibnamefont {Nazeeruddin}},\ }\href@noop {} {\bibfield  {journal} {\bibinfo  {journal} {Journal of Photochemistry and Photobiology C: Photochemistry Reviews}\ }\textbf {\bibinfo {volume} {35}},\ \bibinfo {pages} {1} (\bibinfo {year} {2018})}\BibitemShut {NoStop}%
\bibitem [{\citenamefont {Dale}\ and\ \citenamefont {Scarpulla}(2023)}]{dale2023efficiency}%
  \BibitemOpen
  \bibfield  {author} {\bibinfo {author} {\bibfnamefont {P.~J.}\ \bibnamefont {Dale}}\ and\ \bibinfo {author} {\bibfnamefont {M.~A.}\ \bibnamefont {Scarpulla}},\ }\href@noop {} {\bibfield  {journal} {\bibinfo  {journal} {Solar Energy Materials and Solar Cells}\ }\textbf {\bibinfo {volume} {251}},\ \bibinfo {pages} {112097} (\bibinfo {year} {2023})}\BibitemShut {NoStop}%
\bibitem [{\citenamefont {Li}\ \emph {et~al.}(2025)\citenamefont {Li}, \citenamefont {Luo}, \citenamefont {Wang}, \citenamefont {Li}, \citenamefont {Li}, \citenamefont {Huang}, \citenamefont {Jin}, \citenamefont {Yang}, \citenamefont {Li}, \citenamefont {Zhang} \emph {et~al.}}]{li2025tin}%
  \BibitemOpen
  \bibfield  {author} {\bibinfo {author} {\bibfnamefont {T.}~\bibnamefont {Li}}, \bibinfo {author} {\bibfnamefont {X.}~\bibnamefont {Luo}}, \bibinfo {author} {\bibfnamefont {P.}~\bibnamefont {Wang}}, \bibinfo {author} {\bibfnamefont {Z.}~\bibnamefont {Li}}, \bibinfo {author} {\bibfnamefont {Y.}~\bibnamefont {Li}}, \bibinfo {author} {\bibfnamefont {J.}~\bibnamefont {Huang}}, \bibinfo {author} {\bibfnamefont {Z.}~\bibnamefont {Jin}}, \bibinfo {author} {\bibfnamefont {Y.}~\bibnamefont {Yang}}, \bibinfo {author} {\bibfnamefont {B.}~\bibnamefont {Li}}, \bibinfo {author} {\bibfnamefont {W.}~\bibnamefont {Zhang}}, \emph {et~al.},\ }\href@noop {} {\bibfield  {journal} {\bibinfo  {journal} {Nature}\ ,\ \bibinfo {pages} {1}} (\bibinfo {year} {2025})}\BibitemShut {NoStop}%
\bibitem [{\citenamefont {Frost}\ and\ \citenamefont {Walsh}(2016)}]{frost2016moving}%
  \BibitemOpen
  \bibfield  {author} {\bibinfo {author} {\bibfnamefont {J.~M.}\ \bibnamefont {Frost}}\ and\ \bibinfo {author} {\bibfnamefont {A.}~\bibnamefont {Walsh}},\ }\href@noop {} {\bibfield  {journal} {\bibinfo  {journal} {Accounts of chemical research}\ }\textbf {\bibinfo {volume} {49}},\ \bibinfo {pages} {528} (\bibinfo {year} {2016})}\BibitemShut {NoStop}%
\bibitem [{\citenamefont {Carignano}\ \emph {et~al.}(2017)\citenamefont {Carignano}, \citenamefont {Aravindh}, \citenamefont {Roqan}, \citenamefont {Even},\ and\ \citenamefont {Katan}}]{carignano2017critical}%
  \BibitemOpen
  \bibfield  {author} {\bibinfo {author} {\bibfnamefont {M.~A.}\ \bibnamefont {Carignano}}, \bibinfo {author} {\bibfnamefont {S.~A.}\ \bibnamefont {Aravindh}}, \bibinfo {author} {\bibfnamefont {I.~S.}\ \bibnamefont {Roqan}}, \bibinfo {author} {\bibfnamefont {J.}~\bibnamefont {Even}},\ and\ \bibinfo {author} {\bibfnamefont {C.}~\bibnamefont {Katan}},\ }\href@noop {} {\bibfield  {journal} {\bibinfo  {journal} {The Journal of Physical Chemistry C}\ }\textbf {\bibinfo {volume} {121}},\ \bibinfo {pages} {20729} (\bibinfo {year} {2017})}\BibitemShut {NoStop}%
\bibitem [{\citenamefont {Jinnouchi}\ \emph {et~al.}(2019)\citenamefont {Jinnouchi}, \citenamefont {Lahnsteiner}, \citenamefont {Karsai}, \citenamefont {Kresse},\ and\ \citenamefont {Bokdam}}]{jinnouchi2019phase}%
  \BibitemOpen
  \bibfield  {author} {\bibinfo {author} {\bibfnamefont {R.}~\bibnamefont {Jinnouchi}}, \bibinfo {author} {\bibfnamefont {J.}~\bibnamefont {Lahnsteiner}}, \bibinfo {author} {\bibfnamefont {F.}~\bibnamefont {Karsai}}, \bibinfo {author} {\bibfnamefont {G.}~\bibnamefont {Kresse}},\ and\ \bibinfo {author} {\bibfnamefont {M.}~\bibnamefont {Bokdam}},\ }\href@noop {} {\bibfield  {journal} {\bibinfo  {journal} {Physical review letters}\ }\textbf {\bibinfo {volume} {122}},\ \bibinfo {pages} {225701} (\bibinfo {year} {2019})}\BibitemShut {NoStop}%
\bibitem [{\citenamefont {Wiktor}\ \emph {et~al.}(2023)\citenamefont {Wiktor}, \citenamefont {Fransson}, \citenamefont {Kubicki},\ and\ \citenamefont {Erhart}}]{wiktor2023quantifying}%
  \BibitemOpen
  \bibfield  {author} {\bibinfo {author} {\bibfnamefont {J.}~\bibnamefont {Wiktor}}, \bibinfo {author} {\bibfnamefont {E.}~\bibnamefont {Fransson}}, \bibinfo {author} {\bibfnamefont {D.}~\bibnamefont {Kubicki}},\ and\ \bibinfo {author} {\bibfnamefont {P.}~\bibnamefont {Erhart}},\ }\href@noop {} {\bibfield  {journal} {\bibinfo  {journal} {arXiv preprint arXiv:2304.07402}\ } (\bibinfo {year} {2023})}\BibitemShut {NoStop}%
\bibitem [{\citenamefont {Yi}\ \emph {et~al.}(2016)\citenamefont {Yi}, \citenamefont {Luo}, \citenamefont {Meloni}, \citenamefont {Boziki}, \citenamefont {Ashari-Astani}, \citenamefont {Gr{\"a}tzel}, \citenamefont {Zakeeruddin}, \citenamefont {R{\"o}thlisberger},\ and\ \citenamefont {Gr{\"a}tzel}}]{yi2016entropic}%
  \BibitemOpen
  \bibfield  {author} {\bibinfo {author} {\bibfnamefont {C.}~\bibnamefont {Yi}}, \bibinfo {author} {\bibfnamefont {J.}~\bibnamefont {Luo}}, \bibinfo {author} {\bibfnamefont {S.}~\bibnamefont {Meloni}}, \bibinfo {author} {\bibfnamefont {A.}~\bibnamefont {Boziki}}, \bibinfo {author} {\bibfnamefont {N.}~\bibnamefont {Ashari-Astani}}, \bibinfo {author} {\bibfnamefont {C.}~\bibnamefont {Gr{\"a}tzel}}, \bibinfo {author} {\bibfnamefont {S.~M.}\ \bibnamefont {Zakeeruddin}}, \bibinfo {author} {\bibfnamefont {U.}~\bibnamefont {R{\"o}thlisberger}},\ and\ \bibinfo {author} {\bibfnamefont {M.}~\bibnamefont {Gr{\"a}tzel}},\ }\href@noop {} {\bibfield  {journal} {\bibinfo  {journal} {Energy \& Environmental Science}\ }\textbf {\bibinfo {volume} {9}},\ \bibinfo {pages} {656} (\bibinfo {year} {2016})}\BibitemShut {NoStop}%
\bibitem [{\citenamefont {Beal}\ \emph {et~al.}(2016)\citenamefont {Beal}, \citenamefont {Slotcavage}, \citenamefont {Leijtens}, \citenamefont {Bowring}, \citenamefont {Belisle}, \citenamefont {Nguyen}, \citenamefont {Burkhard}, \citenamefont {Hoke},\ and\ \citenamefont {McGehee}}]{beal2016cesium}%
  \BibitemOpen
  \bibfield  {author} {\bibinfo {author} {\bibfnamefont {R.~E.}\ \bibnamefont {Beal}}, \bibinfo {author} {\bibfnamefont {D.~J.}\ \bibnamefont {Slotcavage}}, \bibinfo {author} {\bibfnamefont {T.}~\bibnamefont {Leijtens}}, \bibinfo {author} {\bibfnamefont {A.~R.}\ \bibnamefont {Bowring}}, \bibinfo {author} {\bibfnamefont {R.~A.}\ \bibnamefont {Belisle}}, \bibinfo {author} {\bibfnamefont {W.~H.}\ \bibnamefont {Nguyen}}, \bibinfo {author} {\bibfnamefont {G.~F.}\ \bibnamefont {Burkhard}}, \bibinfo {author} {\bibfnamefont {E.~T.}\ \bibnamefont {Hoke}},\ and\ \bibinfo {author} {\bibfnamefont {M.~D.}\ \bibnamefont {McGehee}},\ }\href@noop {} {\bibfield  {journal} {\bibinfo  {journal} {The journal of physical chemistry letters}\ }\textbf {\bibinfo {volume} {7}},\ \bibinfo {pages} {746} (\bibinfo {year} {2016})}\BibitemShut {NoStop}%
\bibitem [{\citenamefont {Raval}\ \emph {et~al.}(2022)\citenamefont {Raval}, \citenamefont {Kennard}, \citenamefont {Vasileiadou}, \citenamefont {Dahlman}, \citenamefont {Spanopoulos}, \citenamefont {Chabinyc}, \citenamefont {Kanatzidis},\ and\ \citenamefont {Manjunatha~Reddy}}]{raval2022understanding}%
  \BibitemOpen
  \bibfield  {author} {\bibinfo {author} {\bibfnamefont {P.}~\bibnamefont {Raval}}, \bibinfo {author} {\bibfnamefont {R.~M.}\ \bibnamefont {Kennard}}, \bibinfo {author} {\bibfnamefont {E.~S.}\ \bibnamefont {Vasileiadou}}, \bibinfo {author} {\bibfnamefont {C.~J.}\ \bibnamefont {Dahlman}}, \bibinfo {author} {\bibfnamefont {I.}~\bibnamefont {Spanopoulos}}, \bibinfo {author} {\bibfnamefont {M.~L.}\ \bibnamefont {Chabinyc}}, \bibinfo {author} {\bibfnamefont {M.}~\bibnamefont {Kanatzidis}},\ and\ \bibinfo {author} {\bibfnamefont {G.}~\bibnamefont {Manjunatha~Reddy}},\ }\href@noop {} {\bibfield  {journal} {\bibinfo  {journal} {ACS Energy Letters}\ }\textbf {\bibinfo {volume} {7}},\ \bibinfo {pages} {1534} (\bibinfo {year} {2022})}\BibitemShut {NoStop}%
\bibitem [{\citenamefont {Ren}\ \emph {et~al.}(2022)\citenamefont {Ren}, \citenamefont {Qian}, \citenamefont {Chen}, \citenamefont {Wang},\ and\ \citenamefont {Zhao}}]{ren2022potential}%
  \BibitemOpen
  \bibfield  {author} {\bibinfo {author} {\bibfnamefont {M.}~\bibnamefont {Ren}}, \bibinfo {author} {\bibfnamefont {X.}~\bibnamefont {Qian}}, \bibinfo {author} {\bibfnamefont {Y.}~\bibnamefont {Chen}}, \bibinfo {author} {\bibfnamefont {T.}~\bibnamefont {Wang}},\ and\ \bibinfo {author} {\bibfnamefont {Y.}~\bibnamefont {Zhao}},\ }\href@noop {} {\bibfield  {journal} {\bibinfo  {journal} {Journal of Hazardous Materials}\ }\textbf {\bibinfo {volume} {426}},\ \bibinfo {pages} {127848} (\bibinfo {year} {2022})}\BibitemShut {NoStop}%
\bibitem [{\citenamefont {Hoefler}\ \emph {et~al.}(2017)\citenamefont {Hoefler}, \citenamefont {Trimmel},\ and\ \citenamefont {Rath}}]{hoefler2017progress}%
  \BibitemOpen
  \bibfield  {author} {\bibinfo {author} {\bibfnamefont {S.~F.}\ \bibnamefont {Hoefler}}, \bibinfo {author} {\bibfnamefont {G.}~\bibnamefont {Trimmel}},\ and\ \bibinfo {author} {\bibfnamefont {T.}~\bibnamefont {Rath}},\ }\href@noop {} {\bibfield  {journal} {\bibinfo  {journal} {Monatshefte f{\"u}r Chemie-Chemical Monthly}\ }\textbf {\bibinfo {volume} {148}},\ \bibinfo {pages} {795} (\bibinfo {year} {2017})}\BibitemShut {NoStop}%
\bibitem [{\citenamefont {Jena}\ \emph {et~al.}(2019)\citenamefont {Jena}, \citenamefont {Kulkarni},\ and\ \citenamefont {Miyasaka}}]{jena2019halide}%
  \BibitemOpen
  \bibfield  {author} {\bibinfo {author} {\bibfnamefont {A.~K.}\ \bibnamefont {Jena}}, \bibinfo {author} {\bibfnamefont {A.}~\bibnamefont {Kulkarni}},\ and\ \bibinfo {author} {\bibfnamefont {T.}~\bibnamefont {Miyasaka}},\ }\href@noop {} {\bibfield  {journal} {\bibinfo  {journal} {Chemical reviews}\ }\textbf {\bibinfo {volume} {119}},\ \bibinfo {pages} {3036} (\bibinfo {year} {2019})}\BibitemShut {NoStop}%
\bibitem [{\citenamefont {Ke}\ \emph {et~al.}(2019)\citenamefont {Ke}, \citenamefont {Stoumpos},\ and\ \citenamefont {Kanatzidis}}]{ke2019unleaded}%
  \BibitemOpen
  \bibfield  {author} {\bibinfo {author} {\bibfnamefont {W.}~\bibnamefont {Ke}}, \bibinfo {author} {\bibfnamefont {C.~C.}\ \bibnamefont {Stoumpos}},\ and\ \bibinfo {author} {\bibfnamefont {M.~G.}\ \bibnamefont {Kanatzidis}},\ }\href@noop {} {\bibfield  {journal} {\bibinfo  {journal} {Advanced Materials}\ }\textbf {\bibinfo {volume} {31}},\ \bibinfo {pages} {1803230} (\bibinfo {year} {2019})}\BibitemShut {NoStop}%
\bibitem [{\citenamefont {Aktas}\ \emph {et~al.}(2022)\citenamefont {Aktas}, \citenamefont {Rajamanickam}, \citenamefont {Pascual}, \citenamefont {Hu}, \citenamefont {Aldamasy}, \citenamefont {Di~Girolamo}, \citenamefont {Li}, \citenamefont {Nasti}, \citenamefont {Mart{\'\i}nez-Ferrero}, \citenamefont {Wakamiya} \emph {et~al.}}]{aktas2022challenges}%
  \BibitemOpen
  \bibfield  {author} {\bibinfo {author} {\bibfnamefont {E.}~\bibnamefont {Aktas}}, \bibinfo {author} {\bibfnamefont {N.}~\bibnamefont {Rajamanickam}}, \bibinfo {author} {\bibfnamefont {J.}~\bibnamefont {Pascual}}, \bibinfo {author} {\bibfnamefont {S.}~\bibnamefont {Hu}}, \bibinfo {author} {\bibfnamefont {M.~H.}\ \bibnamefont {Aldamasy}}, \bibinfo {author} {\bibfnamefont {D.}~\bibnamefont {Di~Girolamo}}, \bibinfo {author} {\bibfnamefont {W.}~\bibnamefont {Li}}, \bibinfo {author} {\bibfnamefont {G.}~\bibnamefont {Nasti}}, \bibinfo {author} {\bibfnamefont {E.}~\bibnamefont {Mart{\'\i}nez-Ferrero}}, \bibinfo {author} {\bibfnamefont {A.}~\bibnamefont {Wakamiya}}, \emph {et~al.},\ }\href@noop {} {\bibfield  {journal} {\bibinfo  {journal} {Communications Materials}\ }\textbf {\bibinfo {volume} {3}},\ \bibinfo {pages} {104} (\bibinfo {year} {2022})}\BibitemShut {NoStop}%
\bibitem [{\citenamefont {Lanzetta}\ \emph {et~al.}(2020)\citenamefont {Lanzetta}, \citenamefont {Aristidou},\ and\ \citenamefont {Haque}}]{lanzetta2020stability}%
  \BibitemOpen
  \bibfield  {author} {\bibinfo {author} {\bibfnamefont {L.}~\bibnamefont {Lanzetta}}, \bibinfo {author} {\bibfnamefont {N.}~\bibnamefont {Aristidou}},\ and\ \bibinfo {author} {\bibfnamefont {S.~A.}\ \bibnamefont {Haque}},\ }\href@noop {} {\bibfield  {journal} {\bibinfo  {journal} {The Journal of Physical Chemistry Letters}\ }\textbf {\bibinfo {volume} {11}},\ \bibinfo {pages} {574} (\bibinfo {year} {2020})}\BibitemShut {NoStop}%
\bibitem [{\citenamefont {Dalpian}\ \emph {et~al.}(2017)\citenamefont {Dalpian}, \citenamefont {Liu}, \citenamefont {Stoumpos}, \citenamefont {Douvalis}, \citenamefont {Balasubramanian}, \citenamefont {Kanatzidis},\ and\ \citenamefont {Zunger}}]{dalpian2017changes}%
  \BibitemOpen
  \bibfield  {author} {\bibinfo {author} {\bibfnamefont {G.~M.}\ \bibnamefont {Dalpian}}, \bibinfo {author} {\bibfnamefont {Q.}~\bibnamefont {Liu}}, \bibinfo {author} {\bibfnamefont {C.~C.}\ \bibnamefont {Stoumpos}}, \bibinfo {author} {\bibfnamefont {A.~P.}\ \bibnamefont {Douvalis}}, \bibinfo {author} {\bibfnamefont {M.}~\bibnamefont {Balasubramanian}}, \bibinfo {author} {\bibfnamefont {M.~G.}\ \bibnamefont {Kanatzidis}},\ and\ \bibinfo {author} {\bibfnamefont {A.}~\bibnamefont {Zunger}},\ }\href@noop {} {\bibfield  {journal} {\bibinfo  {journal} {Physical Review Materials}\ }\textbf {\bibinfo {volume} {1}},\ \bibinfo {pages} {025401} (\bibinfo {year} {2017})}\BibitemShut {NoStop}%
\bibitem [{\citenamefont {Kohn}\ and\ \citenamefont {Sham}(1965)}]{kohn1965self}%
  \BibitemOpen
  \bibfield  {author} {\bibinfo {author} {\bibfnamefont {W.}~\bibnamefont {Kohn}}\ and\ \bibinfo {author} {\bibfnamefont {L.~J.}\ \bibnamefont {Sham}},\ }\href@noop {} {\bibfield  {journal} {\bibinfo  {journal} {Physical review}\ }\textbf {\bibinfo {volume} {140}},\ \bibinfo {pages} {A1133} (\bibinfo {year} {1965})}\BibitemShut {NoStop}%
\bibitem [{\citenamefont {Kresse}\ and\ \citenamefont {Hafner}(1993)}]{kresse1993ab}%
  \BibitemOpen
  \bibfield  {author} {\bibinfo {author} {\bibfnamefont {G.}~\bibnamefont {Kresse}}\ and\ \bibinfo {author} {\bibfnamefont {J.}~\bibnamefont {Hafner}},\ }\href@noop {} {\bibfield  {journal} {\bibinfo  {journal} {Physical review B}\ }\textbf {\bibinfo {volume} {47}},\ \bibinfo {pages} {558} (\bibinfo {year} {1993})}\BibitemShut {NoStop}%
\bibitem [{\citenamefont {Giannozzi}\ \emph {et~al.}(2020)\citenamefont {Giannozzi}, \citenamefont {Baseggio}, \citenamefont {Bonf{\`a}}, \citenamefont {Brunato}, \citenamefont {Car}, \citenamefont {Carnimeo}, \citenamefont {Cavazzoni}, \citenamefont {De~Gironcoli}, \citenamefont {Delugas}, \citenamefont {Ferrari~Ruffino} \emph {et~al.}}]{giannozzi2020quantum}%
  \BibitemOpen
  \bibfield  {author} {\bibinfo {author} {\bibfnamefont {P.}~\bibnamefont {Giannozzi}}, \bibinfo {author} {\bibfnamefont {O.}~\bibnamefont {Baseggio}}, \bibinfo {author} {\bibfnamefont {P.}~\bibnamefont {Bonf{\`a}}}, \bibinfo {author} {\bibfnamefont {D.}~\bibnamefont {Brunato}}, \bibinfo {author} {\bibfnamefont {R.}~\bibnamefont {Car}}, \bibinfo {author} {\bibfnamefont {I.}~\bibnamefont {Carnimeo}}, \bibinfo {author} {\bibfnamefont {C.}~\bibnamefont {Cavazzoni}}, \bibinfo {author} {\bibfnamefont {S.}~\bibnamefont {De~Gironcoli}}, \bibinfo {author} {\bibfnamefont {P.}~\bibnamefont {Delugas}}, \bibinfo {author} {\bibfnamefont {F.}~\bibnamefont {Ferrari~Ruffino}}, \emph {et~al.},\ }\href@noop {} {\bibfield  {journal} {\bibinfo  {journal} {The Journal of chemical physics}\ }\textbf {\bibinfo {volume} {152}} (\bibinfo {year} {2020})}\BibitemShut {NoStop}%
\bibitem [{\citenamefont {K{\"u}hne}\ \emph {et~al.}(2020)\citenamefont {K{\"u}hne}, \citenamefont {Iannuzzi}, \citenamefont {Del~Ben}, \citenamefont {Rybkin}, \citenamefont {Seewald}, \citenamefont {Stein}, \citenamefont {Laino}, \citenamefont {Khaliullin}, \citenamefont {Sch{\"u}tt}, \citenamefont {Schiffmann} \emph {et~al.}}]{kuhne2020cp2k}%
  \BibitemOpen
  \bibfield  {author} {\bibinfo {author} {\bibfnamefont {T.~D.}\ \bibnamefont {K{\"u}hne}}, \bibinfo {author} {\bibfnamefont {M.}~\bibnamefont {Iannuzzi}}, \bibinfo {author} {\bibfnamefont {M.}~\bibnamefont {Del~Ben}}, \bibinfo {author} {\bibfnamefont {V.~V.}\ \bibnamefont {Rybkin}}, \bibinfo {author} {\bibfnamefont {P.}~\bibnamefont {Seewald}}, \bibinfo {author} {\bibfnamefont {F.}~\bibnamefont {Stein}}, \bibinfo {author} {\bibfnamefont {T.}~\bibnamefont {Laino}}, \bibinfo {author} {\bibfnamefont {R.~Z.}\ \bibnamefont {Khaliullin}}, \bibinfo {author} {\bibfnamefont {O.}~\bibnamefont {Sch{\"u}tt}}, \bibinfo {author} {\bibfnamefont {F.}~\bibnamefont {Schiffmann}}, \emph {et~al.},\ }\href@noop {} {\bibfield  {journal} {\bibinfo  {journal} {The Journal of Chemical Physics}\ }\textbf {\bibinfo {volume} {152}} (\bibinfo {year} {2020})}\BibitemShut {NoStop}%
\bibitem [{\citenamefont {Fransson}\ \emph {et~al.}(2023)\citenamefont {Fransson}, \citenamefont {Rahm}, \citenamefont {Wiktor},\ and\ \citenamefont {Erhart}}]{fransson2023revealing}%
  \BibitemOpen
  \bibfield  {author} {\bibinfo {author} {\bibfnamefont {E.}~\bibnamefont {Fransson}}, \bibinfo {author} {\bibfnamefont {J.~M.}\ \bibnamefont {Rahm}}, \bibinfo {author} {\bibfnamefont {J.}~\bibnamefont {Wiktor}},\ and\ \bibinfo {author} {\bibfnamefont {P.}~\bibnamefont {Erhart}},\ }\href@noop {} {\bibfield  {journal} {\bibinfo  {journal} {Chemistry of Materials}\ }\textbf {\bibinfo {volume} {35}},\ \bibinfo {pages} {8229} (\bibinfo {year} {2023})}\BibitemShut {NoStop}%
\bibitem [{\citenamefont {Farigliano}\ \emph {et~al.}(2024)\citenamefont {Farigliano}, \citenamefont {Ribeiro},\ and\ \citenamefont {Dalpian}}]{farigliano2024phase}%
  \BibitemOpen
  \bibfield  {author} {\bibinfo {author} {\bibfnamefont {L.~M.}\ \bibnamefont {Farigliano}}, \bibinfo {author} {\bibfnamefont {F.~N.}\ \bibnamefont {Ribeiro}},\ and\ \bibinfo {author} {\bibfnamefont {G.~M.}\ \bibnamefont {Dalpian}},\ }\href@noop {} {\bibfield  {journal} {\bibinfo  {journal} {Materials Advances}\ }\textbf {\bibinfo {volume} {5}},\ \bibinfo {pages} {5794} (\bibinfo {year} {2024})}\BibitemShut {NoStop}%
\bibitem [{\citenamefont {Zhao}\ \emph {et~al.}(2020)\citenamefont {Zhao}, \citenamefont {Dalpian}, \citenamefont {Wang},\ and\ \citenamefont {Zunger}}]{zhao2020polymorphous}%
  \BibitemOpen
  \bibfield  {author} {\bibinfo {author} {\bibfnamefont {X.-G.}\ \bibnamefont {Zhao}}, \bibinfo {author} {\bibfnamefont {G.~M.}\ \bibnamefont {Dalpian}}, \bibinfo {author} {\bibfnamefont {Z.}~\bibnamefont {Wang}},\ and\ \bibinfo {author} {\bibfnamefont {A.}~\bibnamefont {Zunger}},\ }\href@noop {} {\bibfield  {journal} {\bibinfo  {journal} {Physical Review B}\ }\textbf {\bibinfo {volume} {101}},\ \bibinfo {pages} {155137} (\bibinfo {year} {2020})}\BibitemShut {NoStop}%
\bibitem [{\citenamefont {Kaiser}\ \emph {et~al.}(2021)\citenamefont {Kaiser}, \citenamefont {Mosconi}, \citenamefont {Alothman}, \citenamefont {Meggiolaro}, \citenamefont {Gagliardi},\ and\ \citenamefont {De~Angelis}}]{kaiser2021halide}%
  \BibitemOpen
  \bibfield  {author} {\bibinfo {author} {\bibfnamefont {W.}~\bibnamefont {Kaiser}}, \bibinfo {author} {\bibfnamefont {E.}~\bibnamefont {Mosconi}}, \bibinfo {author} {\bibfnamefont {A.~A.}\ \bibnamefont {Alothman}}, \bibinfo {author} {\bibfnamefont {D.}~\bibnamefont {Meggiolaro}}, \bibinfo {author} {\bibfnamefont {A.}~\bibnamefont {Gagliardi}},\ and\ \bibinfo {author} {\bibfnamefont {F.}~\bibnamefont {De~Angelis}},\ }\href@noop {} {\bibfield  {journal} {\bibinfo  {journal} {Materials Advances}\ }\textbf {\bibinfo {volume} {2}},\ \bibinfo {pages} {3915} (\bibinfo {year} {2021})}\BibitemShut {NoStop}%
\bibitem [{\citenamefont {Kaiser}\ \emph {et~al.}(2022)\citenamefont {Kaiser}, \citenamefont {Ricciarelli}, \citenamefont {Mosconi}, \citenamefont {Alothman}, \citenamefont {Ambrosio},\ and\ \citenamefont {De~Angelis}}]{kaiser2022stability}%
  \BibitemOpen
  \bibfield  {author} {\bibinfo {author} {\bibfnamefont {W.}~\bibnamefont {Kaiser}}, \bibinfo {author} {\bibfnamefont {D.}~\bibnamefont {Ricciarelli}}, \bibinfo {author} {\bibfnamefont {E.}~\bibnamefont {Mosconi}}, \bibinfo {author} {\bibfnamefont {A.~A.}\ \bibnamefont {Alothman}}, \bibinfo {author} {\bibfnamefont {F.}~\bibnamefont {Ambrosio}},\ and\ \bibinfo {author} {\bibfnamefont {F.}~\bibnamefont {De~Angelis}},\ }\href@noop {} {\bibfield  {journal} {\bibinfo  {journal} {The Journal of Physical Chemistry Letters}\ }\textbf {\bibinfo {volume} {13}},\ \bibinfo {pages} {2321} (\bibinfo {year} {2022})}\BibitemShut {NoStop}%
\bibitem [{\citenamefont {Mosconi}\ \emph {et~al.}(2015)\citenamefont {Mosconi}, \citenamefont {Azpiroz},\ and\ \citenamefont {De~Angelis}}]{mosconi2015ab}%
  \BibitemOpen
  \bibfield  {author} {\bibinfo {author} {\bibfnamefont {E.}~\bibnamefont {Mosconi}}, \bibinfo {author} {\bibfnamefont {J.~M.}\ \bibnamefont {Azpiroz}},\ and\ \bibinfo {author} {\bibfnamefont {F.}~\bibnamefont {De~Angelis}},\ }\href@noop {} {\bibfield  {journal} {\bibinfo  {journal} {Chemistry of materials}\ }\textbf {\bibinfo {volume} {27}},\ \bibinfo {pages} {4885} (\bibinfo {year} {2015})}\BibitemShut {NoStop}%
\bibitem [{\citenamefont {Carignano}\ \emph {et~al.}(2015)\citenamefont {Carignano}, \citenamefont {Kachmar},\ and\ \citenamefont {Hutter}}]{carignano2015thermal}%
  \BibitemOpen
  \bibfield  {author} {\bibinfo {author} {\bibfnamefont {M.~A.}\ \bibnamefont {Carignano}}, \bibinfo {author} {\bibfnamefont {A.}~\bibnamefont {Kachmar}},\ and\ \bibinfo {author} {\bibfnamefont {J.}~\bibnamefont {Hutter}},\ }\href@noop {} {\bibfield  {journal} {\bibinfo  {journal} {The Journal of Physical Chemistry C}\ }\textbf {\bibinfo {volume} {119}},\ \bibinfo {pages} {8991} (\bibinfo {year} {2015})}\BibitemShut {NoStop}%
\bibitem [{\citenamefont {Behler}\ and\ \citenamefont {Parrinello}(2007)}]{behler2007generalized}%
  \BibitemOpen
  \bibfield  {author} {\bibinfo {author} {\bibfnamefont {J.}~\bibnamefont {Behler}}\ and\ \bibinfo {author} {\bibfnamefont {M.}~\bibnamefont {Parrinello}},\ }\href@noop {} {\bibfield  {journal} {\bibinfo  {journal} {Physical review letters}\ }\textbf {\bibinfo {volume} {98}},\ \bibinfo {pages} {146401} (\bibinfo {year} {2007})}\BibitemShut {NoStop}%
\bibitem [{\citenamefont {Bart{\'o}k}\ \emph {et~al.}(2010)\citenamefont {Bart{\'o}k}, \citenamefont {Payne}, \citenamefont {Kondor},\ and\ \citenamefont {Cs{\'a}nyi}}]{bartok2010gaussian}%
  \BibitemOpen
  \bibfield  {author} {\bibinfo {author} {\bibfnamefont {A.~P.}\ \bibnamefont {Bart{\'o}k}}, \bibinfo {author} {\bibfnamefont {M.~C.}\ \bibnamefont {Payne}}, \bibinfo {author} {\bibfnamefont {R.}~\bibnamefont {Kondor}},\ and\ \bibinfo {author} {\bibfnamefont {G.}~\bibnamefont {Cs{\'a}nyi}},\ }\href@noop {} {\bibfield  {journal} {\bibinfo  {journal} {Physical review letters}\ }\textbf {\bibinfo {volume} {104}},\ \bibinfo {pages} {136403} (\bibinfo {year} {2010})}\BibitemShut {NoStop}%
\bibitem [{\citenamefont {Thompson}\ \emph {et~al.}(2015)\citenamefont {Thompson}, \citenamefont {Swiler}, \citenamefont {Trott}, \citenamefont {Foiles},\ and\ \citenamefont {Tucker}}]{thompson2015spectral}%
  \BibitemOpen
  \bibfield  {author} {\bibinfo {author} {\bibfnamefont {A.~P.}\ \bibnamefont {Thompson}}, \bibinfo {author} {\bibfnamefont {L.~P.}\ \bibnamefont {Swiler}}, \bibinfo {author} {\bibfnamefont {C.~R.}\ \bibnamefont {Trott}}, \bibinfo {author} {\bibfnamefont {S.~M.}\ \bibnamefont {Foiles}},\ and\ \bibinfo {author} {\bibfnamefont {G.~J.}\ \bibnamefont {Tucker}},\ }\href@noop {} {\bibfield  {journal} {\bibinfo  {journal} {Journal of Computational Physics}\ }\textbf {\bibinfo {volume} {285}},\ \bibinfo {pages} {316} (\bibinfo {year} {2015})}\BibitemShut {NoStop}%
\bibitem [{\citenamefont {Sch{\"u}tt}\ \emph {et~al.}(2018)\citenamefont {Sch{\"u}tt}, \citenamefont {Sauceda}, \citenamefont {Kindermans}, \citenamefont {Tkatchenko},\ and\ \citenamefont {M{\"u}ller}}]{schutt2018schnet}%
  \BibitemOpen
  \bibfield  {author} {\bibinfo {author} {\bibfnamefont {K.~T.}\ \bibnamefont {Sch{\"u}tt}}, \bibinfo {author} {\bibfnamefont {H.~E.}\ \bibnamefont {Sauceda}}, \bibinfo {author} {\bibfnamefont {P.-J.}\ \bibnamefont {Kindermans}}, \bibinfo {author} {\bibfnamefont {A.}~\bibnamefont {Tkatchenko}},\ and\ \bibinfo {author} {\bibfnamefont {K.-R.}\ \bibnamefont {M{\"u}ller}},\ }\href@noop {} {\bibfield  {journal} {\bibinfo  {journal} {The Journal of Chemical Physics}\ }\textbf {\bibinfo {volume} {148}} (\bibinfo {year} {2018})}\BibitemShut {NoStop}%
\bibitem [{\citenamefont {Deringer}\ \emph {et~al.}(2019)\citenamefont {Deringer}, \citenamefont {Caro},\ and\ \citenamefont {Cs{\'a}nyi}}]{deringer2019machine}%
  \BibitemOpen
  \bibfield  {author} {\bibinfo {author} {\bibfnamefont {V.~L.}\ \bibnamefont {Deringer}}, \bibinfo {author} {\bibfnamefont {M.~A.}\ \bibnamefont {Caro}},\ and\ \bibinfo {author} {\bibfnamefont {G.}~\bibnamefont {Cs{\'a}nyi}},\ }\href@noop {} {\bibfield  {journal} {\bibinfo  {journal} {Advanced Materials}\ }\textbf {\bibinfo {volume} {31}},\ \bibinfo {pages} {1902765} (\bibinfo {year} {2019})}\BibitemShut {NoStop}%
\bibitem [{\citenamefont {Batzner}\ \emph {et~al.}(2022)\citenamefont {Batzner}, \citenamefont {Musaelian}, \citenamefont {Sun}, \citenamefont {Geiger}, \citenamefont {Mailoa}, \citenamefont {Kornbluth}, \citenamefont {Molinari}, \citenamefont {Smidt},\ and\ \citenamefont {Kozinsky}}]{batzner20223}%
  \BibitemOpen
  \bibfield  {author} {\bibinfo {author} {\bibfnamefont {S.}~\bibnamefont {Batzner}}, \bibinfo {author} {\bibfnamefont {A.}~\bibnamefont {Musaelian}}, \bibinfo {author} {\bibfnamefont {L.}~\bibnamefont {Sun}}, \bibinfo {author} {\bibfnamefont {M.}~\bibnamefont {Geiger}}, \bibinfo {author} {\bibfnamefont {J.~P.}\ \bibnamefont {Mailoa}}, \bibinfo {author} {\bibfnamefont {M.}~\bibnamefont {Kornbluth}}, \bibinfo {author} {\bibfnamefont {N.}~\bibnamefont {Molinari}}, \bibinfo {author} {\bibfnamefont {T.~E.}\ \bibnamefont {Smidt}},\ and\ \bibinfo {author} {\bibfnamefont {B.}~\bibnamefont {Kozinsky}},\ }\href@noop {} {\bibfield  {journal} {\bibinfo  {journal} {Nature communications}\ }\textbf {\bibinfo {volume} {13}},\ \bibinfo {pages} {2453} (\bibinfo {year} {2022})}\BibitemShut {NoStop}%
\bibitem [{\citenamefont {Ko}\ and\ \citenamefont {Ong}(2023)}]{ko2023recent}%
  \BibitemOpen
  \bibfield  {author} {\bibinfo {author} {\bibfnamefont {T.~W.}\ \bibnamefont {Ko}}\ and\ \bibinfo {author} {\bibfnamefont {S.~P.}\ \bibnamefont {Ong}},\ }\href@noop {} {\bibfield  {journal} {\bibinfo  {journal} {Nature Computational Science}\ }\textbf {\bibinfo {volume} {3}},\ \bibinfo {pages} {998} (\bibinfo {year} {2023})}\BibitemShut {NoStop}%
\bibitem [{\citenamefont {Batatia}\ \emph {et~al.}(2023)\citenamefont {Batatia}, \citenamefont {Benner}, \citenamefont {Chiang}, \citenamefont {Elena}, \citenamefont {Kov{\'a}cs}, \citenamefont {Riebesell}, \citenamefont {Advincula}, \citenamefont {Asta}, \citenamefont {Avaylon}, \citenamefont {Baldwin} \emph {et~al.}}]{batatia2023foundation}%
  \BibitemOpen
  \bibfield  {author} {\bibinfo {author} {\bibfnamefont {I.}~\bibnamefont {Batatia}}, \bibinfo {author} {\bibfnamefont {P.}~\bibnamefont {Benner}}, \bibinfo {author} {\bibfnamefont {Y.}~\bibnamefont {Chiang}}, \bibinfo {author} {\bibfnamefont {A.~M.}\ \bibnamefont {Elena}}, \bibinfo {author} {\bibfnamefont {D.~P.}\ \bibnamefont {Kov{\'a}cs}}, \bibinfo {author} {\bibfnamefont {J.}~\bibnamefont {Riebesell}}, \bibinfo {author} {\bibfnamefont {X.~R.}\ \bibnamefont {Advincula}}, \bibinfo {author} {\bibfnamefont {M.}~\bibnamefont {Asta}}, \bibinfo {author} {\bibfnamefont {M.}~\bibnamefont {Avaylon}}, \bibinfo {author} {\bibfnamefont {W.~J.}\ \bibnamefont {Baldwin}}, \emph {et~al.},\ }\href@noop {} {\bibfield  {journal} {\bibinfo  {journal} {arXiv preprint arXiv:2401.00096}\ } (\bibinfo {year} {2023})}\BibitemShut {NoStop}%
\bibitem [{\citenamefont {Batatia}\ \emph {et~al.}(2025)\citenamefont {Batatia}, \citenamefont {Batzner}, \citenamefont {Kov{\'a}cs}, \citenamefont {Musaelian}, \citenamefont {Simm}, \citenamefont {Drautz}, \citenamefont {Ortner}, \citenamefont {Kozinsky},\ and\ \citenamefont {Cs{\'a}nyi}}]{batatia2025design}%
  \BibitemOpen
  \bibfield  {author} {\bibinfo {author} {\bibfnamefont {I.}~\bibnamefont {Batatia}}, \bibinfo {author} {\bibfnamefont {S.}~\bibnamefont {Batzner}}, \bibinfo {author} {\bibfnamefont {D.~P.}\ \bibnamefont {Kov{\'a}cs}}, \bibinfo {author} {\bibfnamefont {A.}~\bibnamefont {Musaelian}}, \bibinfo {author} {\bibfnamefont {G.~N.}\ \bibnamefont {Simm}}, \bibinfo {author} {\bibfnamefont {R.}~\bibnamefont {Drautz}}, \bibinfo {author} {\bibfnamefont {C.}~\bibnamefont {Ortner}}, \bibinfo {author} {\bibfnamefont {B.}~\bibnamefont {Kozinsky}},\ and\ \bibinfo {author} {\bibfnamefont {G.}~\bibnamefont {Cs{\'a}nyi}},\ }\href@noop {} {\bibfield  {journal} {\bibinfo  {journal} {Nature Machine Intelligence}\ ,\ \bibinfo {pages} {1}} (\bibinfo {year} {2025})}\BibitemShut {NoStop}%
\bibitem [{\citenamefont {Drautz}(2019)}]{drautz2019atomic}%
  \BibitemOpen
  \bibfield  {author} {\bibinfo {author} {\bibfnamefont {R.}~\bibnamefont {Drautz}},\ }\href@noop {} {\bibfield  {journal} {\bibinfo  {journal} {Physical Review B}\ }\textbf {\bibinfo {volume} {99}},\ \bibinfo {pages} {014104} (\bibinfo {year} {2019})}\BibitemShut {NoStop}%
\bibitem [{\citenamefont {Jain}\ \emph {et~al.}(2016)\citenamefont {Jain}, \citenamefont {Shin},\ and\ \citenamefont {Persson}}]{jain2016computational}%
  \BibitemOpen
  \bibfield  {author} {\bibinfo {author} {\bibfnamefont {A.}~\bibnamefont {Jain}}, \bibinfo {author} {\bibfnamefont {Y.}~\bibnamefont {Shin}},\ and\ \bibinfo {author} {\bibfnamefont {K.~A.}\ \bibnamefont {Persson}},\ }\href@noop {} {\bibfield  {journal} {\bibinfo  {journal} {Nature Reviews Materials}\ }\textbf {\bibinfo {volume} {1}},\ \bibinfo {pages} {1} (\bibinfo {year} {2016})}\BibitemShut {NoStop}%
\bibitem [{\citenamefont {Hirotsu}\ \emph {et~al.}(1974)\citenamefont {Hirotsu}, \citenamefont {Harada}, \citenamefont {Iizumi},\ and\ \citenamefont {Gesi}}]{hirotsu1974structural}%
  \BibitemOpen
  \bibfield  {author} {\bibinfo {author} {\bibfnamefont {S.}~\bibnamefont {Hirotsu}}, \bibinfo {author} {\bibfnamefont {J.}~\bibnamefont {Harada}}, \bibinfo {author} {\bibfnamefont {M.}~\bibnamefont {Iizumi}},\ and\ \bibinfo {author} {\bibfnamefont {K.}~\bibnamefont {Gesi}},\ }\href@noop {} {\bibfield  {journal} {\bibinfo  {journal} {Journal of the Physical Society of Japan}\ }\textbf {\bibinfo {volume} {37}},\ \bibinfo {pages} {1393} (\bibinfo {year} {1974})}\BibitemShut {NoStop}%
\bibitem [{\citenamefont {Stoumpos}\ \emph {et~al.}(2013)\citenamefont {Stoumpos}, \citenamefont {Malliakas}, \citenamefont {Peters}, \citenamefont {Liu}, \citenamefont {Sebastian}, \citenamefont {Im}, \citenamefont {Chasapis}, \citenamefont {Wibowo}, \citenamefont {Chung}, \citenamefont {Freeman} \emph {et~al.}}]{stoumpos2013crystal}%
  \BibitemOpen
  \bibfield  {author} {\bibinfo {author} {\bibfnamefont {C.~C.}\ \bibnamefont {Stoumpos}}, \bibinfo {author} {\bibfnamefont {C.~D.}\ \bibnamefont {Malliakas}}, \bibinfo {author} {\bibfnamefont {J.~A.}\ \bibnamefont {Peters}}, \bibinfo {author} {\bibfnamefont {Z.}~\bibnamefont {Liu}}, \bibinfo {author} {\bibfnamefont {M.}~\bibnamefont {Sebastian}}, \bibinfo {author} {\bibfnamefont {J.}~\bibnamefont {Im}}, \bibinfo {author} {\bibfnamefont {T.~C.}\ \bibnamefont {Chasapis}}, \bibinfo {author} {\bibfnamefont {A.~C.}\ \bibnamefont {Wibowo}}, \bibinfo {author} {\bibfnamefont {D.~Y.}\ \bibnamefont {Chung}}, \bibinfo {author} {\bibfnamefont {A.~J.}\ \bibnamefont {Freeman}}, \emph {et~al.},\ }\href@noop {} {\bibfield  {journal} {\bibinfo  {journal} {Crystal growth \& design}\ }\textbf {\bibinfo {volume} {13}},\ \bibinfo {pages} {2722} (\bibinfo {year} {2013})}\BibitemShut {NoStop}%
\bibitem [{\citenamefont {Onoda-Yamamuro}\ \emph {et~al.}(1990)\citenamefont {Onoda-Yamamuro}, \citenamefont {Matsuo},\ and\ \citenamefont {Suga}}]{onoda1990calorimetric}%
  \BibitemOpen
  \bibfield  {author} {\bibinfo {author} {\bibfnamefont {N.}~\bibnamefont {Onoda-Yamamuro}}, \bibinfo {author} {\bibfnamefont {T.}~\bibnamefont {Matsuo}},\ and\ \bibinfo {author} {\bibfnamefont {H.}~\bibnamefont {Suga}},\ }\href@noop {} {\bibfield  {journal} {\bibinfo  {journal} {Journal of Physics and Chemistry of Solids}\ }\textbf {\bibinfo {volume} {51}},\ \bibinfo {pages} {1383} (\bibinfo {year} {1990})}\BibitemShut {NoStop}%
\bibitem [{\citenamefont {Whitfield}\ \emph {et~al.}(2016)\citenamefont {Whitfield}, \citenamefont {Herron}, \citenamefont {Guise}, \citenamefont {Page}, \citenamefont {Cheng}, \citenamefont {Milas},\ and\ \citenamefont {Crawford}}]{whitfield2016structures}%
  \BibitemOpen
  \bibfield  {author} {\bibinfo {author} {\bibfnamefont {P.}~\bibnamefont {Whitfield}}, \bibinfo {author} {\bibfnamefont {N.}~\bibnamefont {Herron}}, \bibinfo {author} {\bibfnamefont {W.}~\bibnamefont {Guise}}, \bibinfo {author} {\bibfnamefont {K.}~\bibnamefont {Page}}, \bibinfo {author} {\bibfnamefont {Y.}~\bibnamefont {Cheng}}, \bibinfo {author} {\bibfnamefont {I.}~\bibnamefont {Milas}},\ and\ \bibinfo {author} {\bibfnamefont {M.}~\bibnamefont {Crawford}},\ }\href@noop {} {\bibfield  {journal} {\bibinfo  {journal} {Scientific reports}\ }\textbf {\bibinfo {volume} {6}},\ \bibinfo {pages} {35685} (\bibinfo {year} {2016})}\BibitemShut {NoStop}%
\bibitem [{\citenamefont {Mao}\ \emph {et~al.}(2025)\citenamefont {Mao}, \citenamefont {He}, \citenamefont {Lin}, \citenamefont {Gupta}, \citenamefont {Postec}, \citenamefont {Lanigan-Atkins}, \citenamefont {Krogstad}, \citenamefont {Pajerowski}, \citenamefont {Hong}, \citenamefont {Williams} \emph {et~al.}}]{mao2025correlated}%
  \BibitemOpen
  \bibfield  {author} {\bibinfo {author} {\bibfnamefont {C.}~\bibnamefont {Mao}}, \bibinfo {author} {\bibfnamefont {X.}~\bibnamefont {He}}, \bibinfo {author} {\bibfnamefont {H.-M.}\ \bibnamefont {Lin}}, \bibinfo {author} {\bibfnamefont {M.~K.}\ \bibnamefont {Gupta}}, \bibinfo {author} {\bibfnamefont {P.}~\bibnamefont {Postec}}, \bibinfo {author} {\bibfnamefont {T.}~\bibnamefont {Lanigan-Atkins}}, \bibinfo {author} {\bibfnamefont {M.}~\bibnamefont {Krogstad}}, \bibinfo {author} {\bibfnamefont {D.~M.}\ \bibnamefont {Pajerowski}}, \bibinfo {author} {\bibfnamefont {T.}~\bibnamefont {Hong}}, \bibinfo {author} {\bibfnamefont {T.~J.}\ \bibnamefont {Williams}}, \emph {et~al.},\ }\href@noop {} {\bibfield  {journal} {\bibinfo  {journal} {Physical Review Materials}\ }\textbf {\bibinfo {volume} {9}},\ \bibinfo {pages} {065401} (\bibinfo {year} {2025})}\BibitemShut {NoStop}%
\bibitem [{\citenamefont {Liang}\ \emph {et~al.}(2025)\citenamefont {Liang}, \citenamefont {Klarbring},\ and\ \citenamefont {Walsh}}]{liang2025phase}%
  \BibitemOpen
  \bibfield  {author} {\bibinfo {author} {\bibfnamefont {X.}~\bibnamefont {Liang}}, \bibinfo {author} {\bibfnamefont {J.}~\bibnamefont {Klarbring}},\ and\ \bibinfo {author} {\bibfnamefont {A.}~\bibnamefont {Walsh}},\ }\href@noop {} {\bibfield  {journal} {\bibinfo  {journal} {Chemistry of Materials}\ } (\bibinfo {year} {2025})}\BibitemShut {NoStop}%
\bibitem [{\citenamefont {Plimpton}(1995)}]{lammps}%
  \BibitemOpen
  \bibfield  {author} {\bibinfo {author} {\bibfnamefont {S.}~\bibnamefont {Plimpton}},\ }\href@noop {} {\bibfield  {journal} {\bibinfo  {journal} {Journal of computational physics}\ }\textbf {\bibinfo {volume} {117}},\ \bibinfo {pages} {1} (\bibinfo {year} {1995})}\BibitemShut {NoStop}%
\bibitem [{\citenamefont {Allen}\ \emph {et~al.}(2017)\citenamefont {Allen}, \citenamefont {Tildesley},\ and\ \citenamefont {Tildesley}}]{allen2017}%
  \BibitemOpen
  \bibfield  {author} {\bibinfo {author} {\bibfnamefont {M.}~\bibnamefont {Allen}}, \bibinfo {author} {\bibfnamefont {D.}~\bibnamefont {Tildesley}},\ and\ \bibinfo {author} {\bibfnamefont {D.}~\bibnamefont {Tildesley}},\ }\href {https://books.google.com.br/books?id=nlExDwAAQBAJ} {\emph {\bibinfo {title} {Computer Simulation of Liquids}}}\ (\bibinfo  {publisher} {Oxford University Press},\ \bibinfo {year} {2017})\BibitemShut {NoStop}%
\bibitem [{\citenamefont {Errington}\ and\ \citenamefont {Debenedetti}(2001)}]{Errington2001}%
  \BibitemOpen
  \bibfield  {author} {\bibinfo {author} {\bibfnamefont {J.~R.}\ \bibnamefont {Errington}}\ and\ \bibinfo {author} {\bibfnamefont {P.~G.}\ \bibnamefont {Debenedetti}},\ }\href@noop {} {\bibfield  {journal} {\bibinfo  {journal} {Nature}\ }\textbf {\bibinfo {volume} {409}},\ \bibinfo {pages} {318} (\bibinfo {year} {2001})}\BibitemShut {NoStop}%
\bibitem [{\citenamefont {Mori}\ and\ \citenamefont {Saito}(1986)}]{mori1986x}%
  \BibitemOpen
  \bibfield  {author} {\bibinfo {author} {\bibfnamefont {M.}~\bibnamefont {Mori}}\ and\ \bibinfo {author} {\bibfnamefont {H.}~\bibnamefont {Saito}},\ }\href@noop {} {\bibfield  {journal} {\bibinfo  {journal} {Journal of Physics C: Solid State Physics}\ }\textbf {\bibinfo {volume} {19}},\ \bibinfo {pages} {2391} (\bibinfo {year} {1986})}\BibitemShut {NoStop}%
\bibitem [{\citenamefont {Fabini}\ \emph {et~al.}(2016)\citenamefont {Fabini}, \citenamefont {Laurita}, \citenamefont {Bechtel}, \citenamefont {Stoumpos}, \citenamefont {Evans}, \citenamefont {Kontos}, \citenamefont {Raptis}, \citenamefont {Falaras}, \citenamefont {Van~der Ven}, \citenamefont {Kanatzidis} \emph {et~al.}}]{fabini2016dynamic}%
  \BibitemOpen
  \bibfield  {author} {\bibinfo {author} {\bibfnamefont {D.~H.}\ \bibnamefont {Fabini}}, \bibinfo {author} {\bibfnamefont {G.}~\bibnamefont {Laurita}}, \bibinfo {author} {\bibfnamefont {J.~S.}\ \bibnamefont {Bechtel}}, \bibinfo {author} {\bibfnamefont {C.~C.}\ \bibnamefont {Stoumpos}}, \bibinfo {author} {\bibfnamefont {H.~A.}\ \bibnamefont {Evans}}, \bibinfo {author} {\bibfnamefont {A.~G.}\ \bibnamefont {Kontos}}, \bibinfo {author} {\bibfnamefont {Y.~S.}\ \bibnamefont {Raptis}}, \bibinfo {author} {\bibfnamefont {P.}~\bibnamefont {Falaras}}, \bibinfo {author} {\bibfnamefont {A.}~\bibnamefont {Van~der Ven}}, \bibinfo {author} {\bibfnamefont {M.~G.}\ \bibnamefont {Kanatzidis}}, \emph {et~al.},\ }\href@noop {} {\bibfield  {journal} {\bibinfo  {journal} {Journal of the American Chemical Society}\ }\textbf {\bibinfo {volume} {138}},\ \bibinfo {pages} {11820} (\bibinfo {year} {2016})}\BibitemShut {NoStop}%
\end{thebibliography}%

\end{document}